 \documentclass[apj]{emulateapj}
\usepackage{color,ulem}
\usepackage{amsmath}

\slugcomment{draft v2}

\shorttitle{BBH spins and progenitors }
\shortauthors{Hotokezaka \& Piran}

\begin{document}


\title{Implications of  the low binary black hole {aligned} spins observed by LIGO}


\author{Kenta Hotokezaka\altaffilmark{1,2}}
\author{Tsvi Piran\altaffilmark{2}}


\altaffiltext{1}{Center for Computational Astrophysics, Flatiron Institute, 162 5th Ave, New York, 10010, NY,  USA}
\altaffiltext{2}{Racah Institute of Physics, The Hebrew University of Jerusalem, Jerusalem 91904, Israel}


\begin{abstract}
We explore the implications of the observed low spins aligned with the orbital axis in Advanced~LIGO~O1 run  on  binary black hole (BBH)~merger  scenarios in which the merging BBHs have evolved from field binaries.
The coalescence time determines the initial orbital separation of BBHs. This, in turn, determines whether the stars are synchronized before collapse and hence  determines their projected spins. 
Short coalescence times   imply  synchronization and large spins.  Among known stellar objects, Wolf-Rayet~(WR) stars seem the only  progenitors consistent with the low aligned spins observed in LIGO's~O1 
provided that the orbital axis maintains its direction during the collapse. 
We calculate the  spin distribution of  BBH mergers in  the local Universe and its redshift evolution for  WR progenitors.  Assuming that the BBH formation rate peaks around a redshift of $\sim 2\,$--$\,3$,  we show that BBH mergers in the local Universe are dominated  by  low spin events. The high spin population starts to dominate at a redshift of $\sim0.5\,$--$\,1.5$. WR stars are also progenitors of long Gamma-Ray~Bursts~(LGRBs) that  take place at a comparable rate to BBH mergers.  We discuss the possible connection between  the two phenomena.  Additionally, we  show that  hypothetical Population~III star progenitors are also possible. 
Although WR and Population~III progenitors are consistent with the current data, both models predict a non-vanishing fraction of high 
aligned spin BBH mergers. If those are not detected within the coming LIGO/Virgo runs,  it will be unlikely that the observed BBHs  formed via field binaries.
\end{abstract}


\keywords{ gravitational wave ---
black hole physics ---
gamma-ray burst: general ---
stars: black holes ---
stars: massive ---}

\section{Introduction}
The Advanced LIGO gravitational-wave (GW) detectors discovered binary black holes (BBHs) mergers. 
\citep{abbott2016PhRvL}.
The discovery has  opened gravitational-wave astronomy of black holes. 
The GW measurements using a matched filter analysis
provide us valuable information on the GW sources, e.g, the masses
and spins of the BBHs. In addition, the luminosity distance or the cosmological redshift
of the sources can also be measured, and thus, 
the event rate of BBH mergers is obtained. The resulting mass function of the primaries
is consistent with the Salpeter initial mass function \citep{abbott2016PhRvX}. 
Additionally, 
the inferred event rate is surprisingly high,  about $0.1\%$ of the current core-collapse supernova rate, 
suggesting that these are not the results of an obscure rare phenomenon. 
These facts motivate us to consider here the formation pathway of merging BBHs and their binary evolution
and the impact on understanding astrophysical phenomena involving
stellar mass black holes (see e.g. \citealt{abbott2016ApJ} and references therein).

The formation pathway of merging BBHs is one of the most intriguing  mysteries 
that arose after the LIGO's discovery.  One of the puzzles is how do so massive BBHs form in close binary systems. 
Such massive stellar progenitors
are expected  to evolve to giant stars whose stellar radii exceed 
significantly the semi-major axis that allows BBHs to merge within a Hubble time.

A possible scenario is one involving a dynamically-unstable common envelope phase (see e.g. \citealt{belczynski2016Nature}).
While  a lot of works has been dedicated to this issue (see, e.g., \citealt{kruckow2016A&A} for a recent work
and \citealt{ivanova2013A&ARv} and references therein), 
the outcome of common envelope phases is unknown. 
Other scenarios that avoid common envelope phases  include  chemically homogeneous
evolution \citep{mandel2016MNRAS},  rapid-mass transfers  \citep{vandenHeuvel2017},
massive overcontact binaries  \citep{marchant2016}, 
and Population (Pop) III progenitors \citep{kinugawa2014MNRAS,inayoshi2017}.
We consider these scenarios here. We don't discuss  here other
 scenarios that are not based on binary stellar evolution: a dynamical capture 
in dense stellar clusters \citep{rodriguez2016ApJb,oleary2016ApJ},
formation in galactic nuclei \citep{antonini2016ApJ,bartos2017ApJ,stone2017MNRAS},
and primordial BBHs \citep{sasaki2016PhRvL,bird2016PhRvL,blinnikov2016JCAP,kashlinsky2016ApJ}.

\begin{table*}[t]
\begin{center}
\caption[]{Parameters of the BBH mergers detected during LIGO's O1 Run}
\label{tab:LIGO}
\scalebox{1.2}{\begin{tabular}{lccccc}
\hline \hline
Event & $m_1\,[M_{\odot}]$ & $m_2\,[M_{\odot}]$  & $m_{\rm tot}\,[M_{\odot}]$ & $\chi_{\rm eff}$ & Rate [${\rm Gpc^{-3}\,yr^{-1}}$]  \\ \hline
GW150914 & $36.2^{+5.2}_{-3.8}$ & $29.1^{+3.7}_{-4.4}$ & $65.3^{+4.1}_{-3.4}$ & $-0.06^{+0.14}_{-0.14}$ & $3.4^{+8.6}_{-2.8}$\\
GW151226 & $14.2^{+8.3}_{-3.7}$ & $7.5^{+2.3}_{-2.3}$ & $21.8^{+5.9}_{-1.7}$ & $0.21^{+0.20}_{-0.10}$ & $37^{+92}_{-31}$\\
LVT151012 & $23^{+18}_{-6}$ & $13^{+4}_{-5}$ & $37^{+13}_{-4}$ & $0.0^{+0.3}_{-0.2}$ & $9.4^{+30.4}_{-8.7}$\\
\hline \hline 
\end{tabular}}
{\\
\hspace{-4.6cm} The parameters are median values with $90\%$ confidence intervals.\\
\hspace{-6.7cm} The values are taken from \cite{abbott2016PhRvX}.
}
\end{center}
\end{table*}

A route to approach the
progenitor scenario, on which we focus on in this paper, is to 
deriving the required conditions for the progenitors of BBH mergers from  the observed parameters of the 
systems.   
This method allows us to avoid numerous uncertainties in modeling of the 
stellar evolution and the binary interaction. \cite{kushnir2016MNRAS} have recently pointed out that among the
observable quantities the spin of merging BBHs {parallel to the orbital axis}
seems to be the most useful to constrain the progenitor properties (see also \citealt{zaldarriaga2017}). They have shown that
the coalescence time of GW150914 is longer than $1$~Gyr,
if this merger arose  from Wolf-Rayet (WR) stars in a field binary system.
{These discussions assume that  natal kicks during the collapse
don't change significantly  the orbital angular momentum so that the aligned spin parameters
are expected to have positive values. It is worthy noting that the aligned spin parameters measured by LIGO can be negative. 
Such negative values  are naturally expected in the
dynamical capture scenario \citep{rodriguez2016ApJ}. }

The event rate of BBH mergers inferred by the LIGO's detections
is similar to the rate of long Gamma-Ray Bursts (LGRBs) after the beaming correction with
a reasonable value \citep{wanderman2010MNRAS}. LGRBs are produced during 
the core collapse of massive stars. They are believed to form by a black hole surrounded by an accretion disk  \citep{woosley1993ApJ} , which  requires rapid rotation of the progenitor.  These facts motivate us to explore the possibility that LGRBs are produced
during the core collapse of massive stars in close binaries which
eventually evolve to merging BBHs. In fact, such scenarios in which 
LGRBs arise from massive stars in close binaries have been
already discussed  (e.g. \citealt{podsiadlowski2004ApJ,detmers2008A&A,woosley2012ApJ}).

In this paper, we consider the spins  of BBH mergers 
for different types of progenitors and
estimate the expected  spin distribution and its redshift distribution.
We briefly summarize the observed aligned spins of the
BBH mergers detected in LIGO's O1 run in \S \ref{sec:LIGO}.
We describe the  spin and tidal synchronization of the progenitors 
in \S \ref{sec:form} and \S \ref{sec:syn} and discuss
different stellar models in \S \ref{sec:model}.
The possible connection between the BBH merger progenitors
and LGRBs is discussed in \S \ref{sec:LGRBs}.
We show the spin distribution and its redshift evolution 
for the case of WR  progenitors and Pop III progenitors in \S \ref{sec:spin}.
We also discuss caveats of the spin argument in \S \ref{sec:discussion}.
We conclude our results in \S \ref{sec:conclusion}.
In this paper, we use a ${\rm \Lambda CDM}$ cosmology
with $h=0.7$, $\Omega_{\Lambda}=0.7$, and $\Omega_{M}=0.3$.

\section{LIGO's O1 GW detections} \label{sec:LIGO}
{\it Mass function and Rate:} The masses and event rates of the three BBHs detected in LIGO's O1 run
are summarized in Table \ref{tab:LIGO}.
 These event rates suggest that the primary mass function of
BBH mergers is $dR/dm_1 \propto m_1^{-\alpha}$, 
where $\alpha=2.5^{+1.5}_{-1.6}$ and $m_1$ is the mass of the primaries. The total BBH merger rate density
is then $99_{-70}^{+138}\,{\rm Gpc^{-3}\,yr^{-1}}$ for $\alpha = 2.35$
and $m_{1,{\rm min}}=5M_{\odot}$,
where this minimal mass is based on the observed population of these mergers  \citep{abbott2016PhRvX}. This choice
 is consistent 
with observations of Galactic black holes (see, e.g, \citealt{ozel2010ApJ,farr2011ApJ}).  
Note that the total event rate is sensitive to the choice of $m_{1,{\rm min}}$ that is still uncertain. If we take the secondary mass of GW151226, $7.5M_{\odot}$, as the minimal black hole mass in BBH mergers,  the total event rate decreases to $57\,{\rm Gpc^{-3}\,yr^{-1}}$.

This primary mass function is consistent with the Salpeter initial mass
function of local stars \citep{abbott2016PhRvX}, suggesting that these BBHs may originate from
binary stellar objects. In addition, the event rate is similar to that
of  LGRBs, which are thought to be associated with 
black hole formations. In \S \ref{sec:spin} and \S \ref{sec:discussion}, we will discuss a scenario motivated by this similarity in which
LGRBs are produced at the core collapse of massive stars in binary systems that eventually evolve to BBH mergers.

{\it Spin parameters:}  The spin angular momentum of the merging BBHs can be inferred from the
  gravitational-wave signals. 
The effective spin parameter $\chi_{\rm eff}$ is  a mass-weighted
mean spin angular momentum of the two black holes
parallel to the orbital angular momentum:
\begin{equation}
\chi_{\rm eff} \equiv  \ \frac{m_1}{m_{\rm tot}}(\vec{s}_1\cdot \hat{L})+\frac{m_2}{m_{\rm tot}}(\vec{s}_2\cdot \hat{L}), 
\end{equation} 
where $m_2$ is the secondary mass, $m_{\rm tot}=m_1 + m_2$, $\vec{s}_1$ and $\vec{s}_2$
are the specific spin angular momenta of the primary and secondary
normalized by the speed of light $c$, gravitational constant $G$, and mass of each component, and $\hat{L}$ is 
the unit vector of the orbital angular momentum.
This is well constrained as compared with the individual
component spins that are not. The measured values are shown in Table \ref{tab:LIGO}.
$-1 \le \chi_{\rm eff}\le 1 $, where the lower limit arises when both black holes' spins are maximal and anti-aligned 
to the orbital axis and the upper limit when both are maximal and aligned. If one of the black holes' spins 
is maximal and aligned and the other one is not we expect $\chi_{\rm eff}\approx 0.5 $ for equal mass BBHs.
 The observed  values of $\chi_{\rm eff}$ clearly
exclude rapidly rotating synchronized progenitors {whose spin axis is parallel to the orbital axis.}
As pointed out by \cite{kushnir2016MNRAS,rodriguez2016ApJ},
these measured {effective} spin parameters depend sensitively on the
evolutional path of progenitors of BBHs and provide important constraints on the origin 
of BBH mergers. We  focus on 
the spin evolution of the BBH progenitors in the rest of
the paper. Note that the error range of the observed $\chi_{\rm eff}$ of GW151226 
does not exclude the possibility that the spin parameter of the secondary 
is of order unity if the primary's spin is much smaller than unity.

\section{Binary black hole progenitors' spin}\label{sec:form}

A binary system with stellar masses  $m_1$ and $m_2$ at a semi-major axis  $a$
inspirals in due to gravitational-wave radiation. The time until the coalescence, $t_c$, is \citep{peters1964PhRv}:
\begin{eqnarray}
t_c & = & \frac{5}{256} \frac{a}{c}\frac{c^2a}{Gm_1}\frac{c^2a}{Gm_2}\frac{c^2a}{Gm_{\rm tot}} 
\\
&\approx& 10q^2\left(\frac{2}{1+q}\right)\left(\frac{a}{44R_{\odot}}\right)^4
\left(\frac{m_2}{30M_{\odot}}\right)^{-3}~{\rm Gyr} , \nonumber 
\end{eqnarray}
where $q\equiv m_2/m_1$. 
For binaries with $t_c=10$~Gyr, the corresponding  orbital period is: 
$P_{\rm orb} 
 \approx  4.4\,{\rm day}\, (a/44R_{\odot})^{2/3}
(m_{\rm tot}/60M_{\odot})^{-1/2}$.
{Note that, for simplicity,  we assume circular orbits here and elsewhere.
Such  orbits are expected within most binary evolution scenarios 
(see, e.g., \citealt{zahn1977A&A} for the orbital circularization due to the tidal torque),
as long as  natal kicks 
at the black hole formation are not significant. }

The stellar radius cannot exceed the Roche limit.
The Roche limit of the secondary \citep{eggleton1983ApJ} is   
$R_{\rm RL}\approx 0.49q^{2/3}a/[0.6q^{2/3}+\ln (1+q^{1/3})]$.
For equal mass binaries: $R_{\rm RL}\approx 0.38a$.
We denote hereafter the primary (secondary) as the first (second) star  evolving to  core collapse. 
Requiring $R_2<R_{\rm RL}$ and  a coalescence time less than a Hubble time 
yields:
\begin{eqnarray}
R_2\lesssim  17 R_{\odot}(m_{2}/30M_{\odot})^{3/4},\label{roche} 
\end{eqnarray}
where $R_2$ is the stellar radius of the secondary and we have assumed $q=1$.
In the rest of the paper, we consider  massive stars that satisfy this condition.

Clearly if the stellar spin just before the collapse  is larger than the maximal Kerr black hole spin, some mass and angular momentum will be 
shed out and the formed black hole will be a maximal Kerr. Otherwise, the spin angular momentum of the black hole 
equals to its progenitor's  one (see, e.g., \citealt{barkov2010MNRAS},
and also \citealt{sekiguchi2011ApJ,oconnor2011ApJ} for numerical studies). 
A critical question is whether the star is synchronized (tidally locked) with the orbital motion before the collapse.  We characterize this by a synchronization parameter $x_{s}$, e.g., $x_s=1$ and $0$
correspond to a  star tidally synchronized with the orbital motion
and a non-rotating star, respectively. We don't expect negative values of $x_s$ (counter rotating stars)
and values larger than unity. 

If  there are no significant mass and angular momentum losses  from the system  during the collapse
the spin of the secondary black hole is characterized by its stellar mass, radius, and semi-major axis:
\begin{equation}
J_2  =  x_s  I_2\Omega_{\rm orb},
 =  x_s \epsilon m_2 R_2^2 \left(\frac{Gm_{\rm tot}}{a^3}\right)^{1/2},
\end{equation}
where 
 $\epsilon$ characterizes the star's moment of inertia  $I_2 \equiv \epsilon m_2 R_2^2$.
 Here and in the following, we consider,  for simplicity,  rigidly rotating stars.
The spin parameter is then
\begin{eqnarray}
\chi_2 & \equiv&  \frac{J_2}{m_2 r_{g,2}c},\label{stellarspin}
   \\
 &= & x_s\epsilon\left(\frac{R_2}{r_{g,2}}\right)^{1/2}
 \left(\frac{R_2}{a}\right)^{3/2}
  \left(\frac{m_{\rm tot}}{m_{2}}\right)^{1/2} \approx 
 x_s\left(\frac{\epsilon}{0.075}\right) \nonumber \\ 
&\times &  \left(\frac{R_2}{4.7R_{\odot}}\right)^2
 \left(\frac{a}{44R_{\odot}}\right)^{-3/2}    
 \left(\frac{m_{\rm tot}}{2m_{2}}\right)^{1/2}   \left(\frac{m_{2}}{30 M_{\odot}}\right)^{-1/2}, 
\nonumber
\end{eqnarray}
where $r_{g,2}\equiv Gm_{2}/c^{2}$. The normalizations of $R_2$ and $a$ were chosen so that 
the spin parameter is unity for $x_s=1$ and the merger takes place on a time scale of 10 Gyr.

The spin parameter can be directly related to the merger time scale:
\begin{eqnarray}
\chi_2 
& \approx & x_s~
q^{1/4}\left(\frac{1+q}{2} \right)^{1/8} \left(\frac{\epsilon}{0.075}\right) \label{spin} \\
& &\times\left(\frac{t_c}{10\,{\rm Gyr}}\right)^{-3/8}  \left(\frac{R_2}{4.7R_{\odot}}\right)^{2}
\left(\frac{m_2}{30M_{\odot}}\right)^{-13/8}. \nonumber
\end{eqnarray}
\\

\section{Synchronization}\label{sec:syn}
In close binary systems, the tidal torque on the stars forces them to
reach an equilibrium state, where the stellar rotation is synchronized with
the orbital motion. The synchronization timescale
of a star with a radiative envelope and a convective core  can be estimated as
\begin{eqnarray}
t_{\rm syn} &\approx&  0.07~{\rm Myr}~q^{-2}\left(\frac{1+q}{2}\right)^{-5/6} 
\left(\frac{\epsilon}{0.075}\right)
\left(\frac{R}{14R_{\odot}}\right)^{-7}  \nonumber \\
& &\times \left(\frac{M}{30M_{\odot}}\right)^{-1/2}
\left(\frac{a}{44R_{\odot}}\right)^{17/2} \left(\frac{E_{2}}{10^{-6}}\right)^{-1},
\label{eq:Zahn}
\end{eqnarray}
where $E_2$ is a dimensionless quantity depending on 
the stellar structure introduced by \cite{zahn1975A&A}. 
$E_2$ is  $\sim 10^{-7}$--$10^{-4}$ for 
massive main sequence stars and WR stars \citep{zahn1975A&A,kushnir2017MNRAS}. 
It may be smaller for blue supergiants.
For WR progenitors, \cite{kushnir2016MNRAS} derive an useful form of Eq. (\ref{eq:Zahn}) 
as:
\begin{eqnarray}
t_{\rm syn} \approx 10\,{\rm Myr}\,q^{-1/8}\left(\frac{1+q}{2q} \right)^{31/24}
\left(\frac{t_c}{1\,{\rm Gyr}} \right)^{17/8}.
\label{eq:kush}
\end{eqnarray}
We will use this form for WR progenitors in \S \ref{sec:spin}.

If the synchronization time is much shorter than other timescales, e.g., the stellar lifetime and
the wind angular-momentum loss timescale,
the star is synchronized with the orbital motion, i.e., $x_s=1$. On the contrary, if
the synchronization time is much longer than the others, the stellar spin parameter
decreases with time due to the wind loss from the initial value.

If the synchronization timescale is comparable to the stellar lifetime or
the wind timescale, which is the case for WR progenitors, one needs to solve the time evolution of the synchronization parameter. 
In the following we estimate the spin evolution of close binary WR stars 
using the formulation of 
\cite{kushnir2016MNRAS},  that takes  into account the tidal synchronization,
wind mass loss, and the stellar lifetime. 
Given an initial value $x_{s,i}$ at the beginning of the WR phase, the 
 synchronization parameter evolves  as:
\begin{eqnarray}
\frac{dx_s}{d\tau} = \frac{t_w}{t_{\rm syn}}(1-x_s)^{8/3}-x_s,
\end{eqnarray} 
where $t_w$ is the time scale of spin angular momentum loss
and $\tau=t/t_w$. The solution approaches to an equilibrium value, $x_{s,{\rm eq}}$, 
at late times:
\begin{eqnarray}
\frac{t_w}{t_{\rm syn}} (1-x_{s,{\rm eq}})^{8/3} = x_{s,{\rm eq}}.\label{xeq}
\end{eqnarray}  
Note, however, that $t$ cannot exceed the stellar lifetime $t_{*}$.
The approximate solutions at $t_*$ are summarized in \cite{kushnir2016MNRAS} for
different parameter regions.

If the timescale of the angular momentum loss due to the wind
is longer than the stellar lifetime, the synchronization parameter of
a star at the end of its lifetime, $x_{s,f}$, can be estimated as:
\begin{equation}
x_{s,f} \approx 
\begin{cases} 
{\rm max} (1-t_{*}/t_w,~x_{s,{\rm eq}}) & {\rm for~~} x_{s,i}=1, \\
{\rm min}(t_w/t_{\rm syn}, x_{s,{\rm eq}}) &  {\rm for~~}x_{s,i}=0.
\end{cases}
\end{equation}


In order to estimate the synchronization parameter of
the WR stars at the end of their life in \S \ref{sec:spin}, we will 
use the above solutions  with the following parameters:
 $t_w=1$~Myr and
 $t_{\rm WR}=0.3$~Myr \citep{langer1994A&A,meynet2003A&A,maynet2005A&A}. 
Using the synchronization parameter we calculate the spin parameter of the 
individual black holes for a given mass, radius, and
coalescence  time $t_c$.

{For the wind angular momentum losses, we assume that  isotropic winds remove 
the outermost spherical shell of a star with spin angular momentum 
of $2M_s R_*^2 \Omega_*/3$,
where $M_s$ is the mass of the shell, $R_*$ and $\Omega_*$ are
the stellar radius and spin angular frequency. In this description, $t_w\approx 1$~Myr
corresponds to a mass loss rate of $\sim 10^{-5.5}M_{\odot}$/yr. 
Note that, however, the efficiency of the spin
angular momentum  loss due to winds can be either lower  or higher.
For instance, winds  may have
anisotropic structures due to a fast stellar rotation, that removes less spin angular momentum \citep{meynet2007A&A}. On the contrary,
magnetic winds carry out spin angular momentum more efficiently
depending on the field strength (e.g. \citealt{Ud-Doula2009MNRAS}). }
Note also that a significant mass loss from a binary 
increases the semi-major axis. However, 
because 
the spin angular momentum is more efficiently lost from the stellar surface 
for rigidly rotating stars, the mass loss timescale is $\sim 10 t_w (0.075/\epsilon)$ for the isotropic winds.
 With the parameters we consider here,
this effect on the semi-major axis is negligible.


{ The gravitational-wave measurements are quite insensitive
to the spin components perpendicular to the orbital axis as the observed $\chi_{\rm eff}$ 
measures only the spin parallel to the orbital axis. 
In the following discussion, 
we assume that the misalignments of the BBHs' spin axes to the orbital axis are negligible, 
and hence,   
the synchronization parameters cannot be negative. This assumption is valid 
as long as the progenitor binaries do not receive  significant kicks at the black hole formation.
As we will discuss in \S \ref{sec:discussion},  such kicks are unlikely when black holes form. 
  On the contrary, the dynamical capture scenario of 
merging BBH formation naturally predicts that a half of  mergers 
have negative values of $\chi_{\rm eff}$. Note that the measured effective spin parameters of GW150914 and LVT151012
allow negative values within errors. 
Therefore the discovery of BBHs with a negative 
$\chi_{\rm eff}$ will have a significant impact on the understanding of the formation scenarios
and the kick at the black hole formation \citep{rodriguez2016ApJ}. 
}

\section{synchronization for different stellar models} \label{sec:model}
As the stellar radius and resulting black hole's  spin are 
tightly connected, the spin measurements strongly 
constrain  the possible progenitors of the observed BBH mergers. 
Population synthesis calculations considering  
the stellar evolution and the binary interactions are often used 
to estimate the rate, mass, and spin distribution of compact binary 
mergers and to discuss their progenitors.  Here 
we take a  different approach. 
We don't discuss the binary evolution. Instead we  focus on the the observed  low {aligned} spins
and examine their implications 
on the stellar progenitor  just before its
 core  collapse to a black hole.

We  consider known types of massive
stellar objects and hypothetical Pop III stars.
The BBH mergers event rate suggests that,  
if they form  via  binary stellar evolution,  there are $\sim 10$ or less such   progenitors
 in the Galaxy (using  $0.01~{\rm Mpc^{-3}}$ as the number density of the Milky-Way size galaxies
and a stellar lifetime of $1$~Myr). 
With such a small number it is possible and even likely 
that we have not identified these objects in the Galaxy. It is interesting to note, in passing, 
that Gaia might be able to identify these binaries as they involve the most massive stars and hence most luminous stars. 
 
\begin{figure}[t]
  \begin{center}
    \includegraphics[width=85mm]{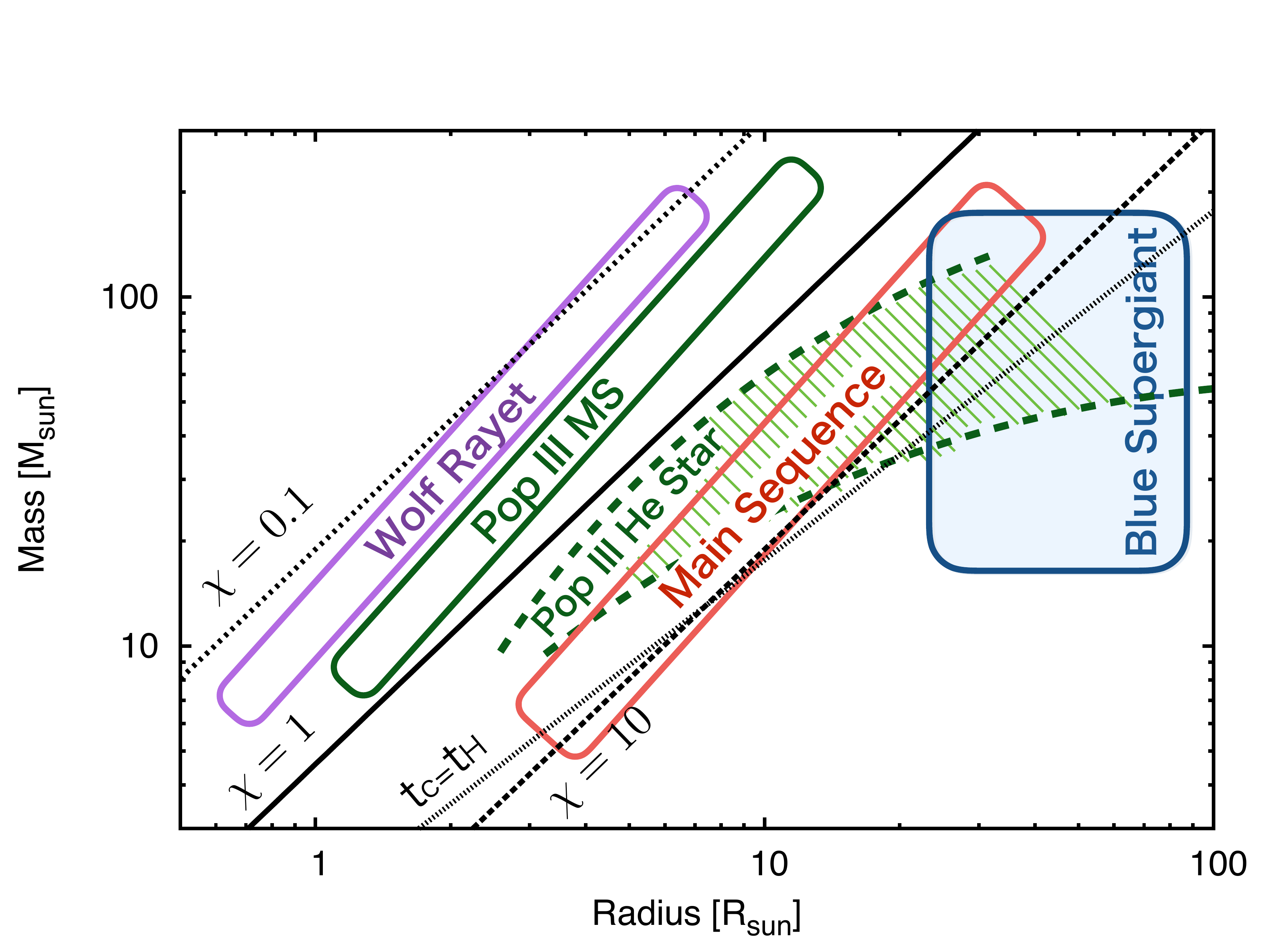}
    \caption{Mass - radius relations of different stellar models.
    The  diagonal (dotted, solid and short-dashed black) lines depict the resulting black hole dimensionless spin for these masses and radii  assuming
    that the star is synchronized at a semi-major axis where the coalescence time
    is a Hubble time. Also shown as a diagonal line labeled by $t_c=t_{\rm H}$ is the stellar radius limited by
    the Roche limit. Stars in the right side of this line cannot exist in a binary system
    whose coalescence time is less than a Hubble time. The curves are drown for a
    mass ratio, q, of unity.  One can clearly see that most models will result in $\chi$ values
    much larger than unity.
}
    \label{fig:MR}
  \end{center}
\end{figure}

\begin{figure*}[t]
  \begin{center}
    \includegraphics[width=70mm]{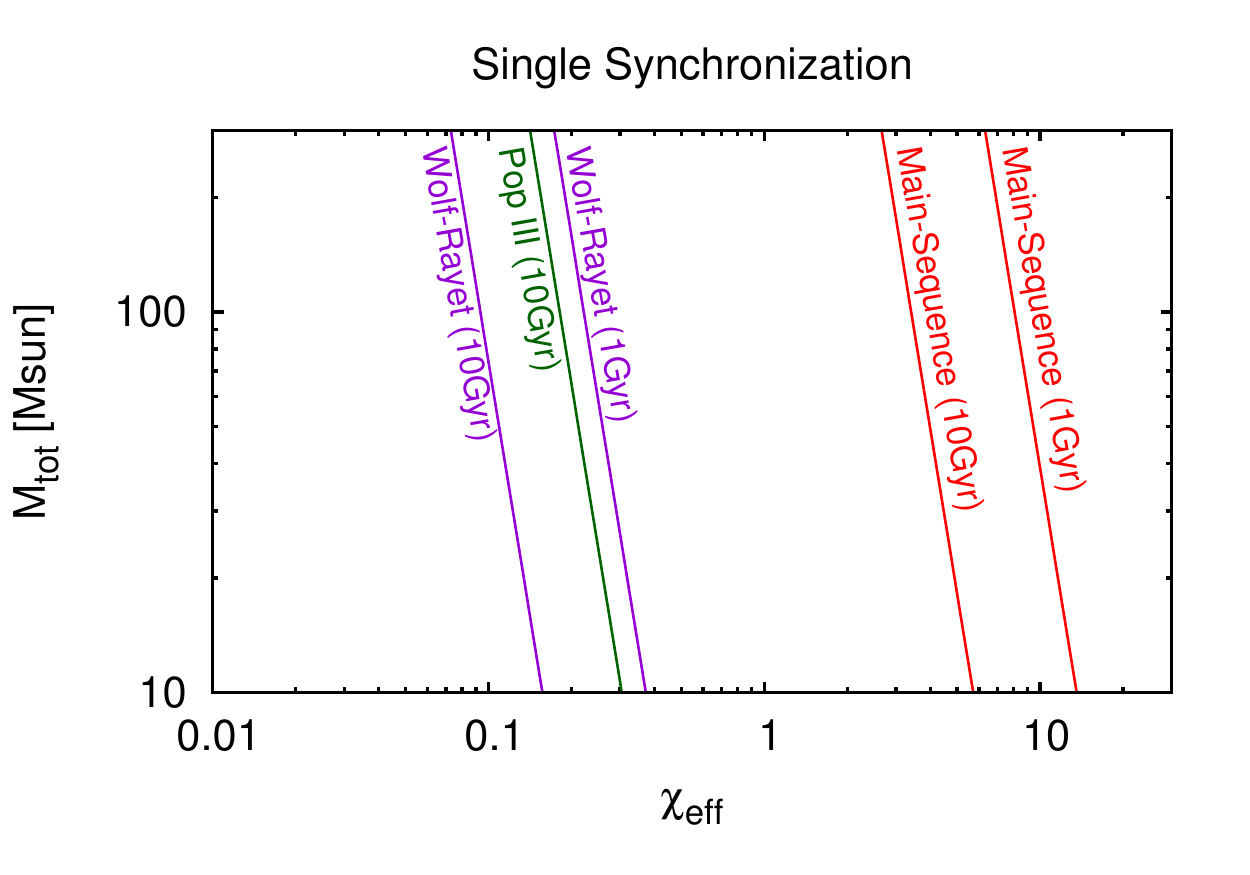}
     \includegraphics[width=70mm]{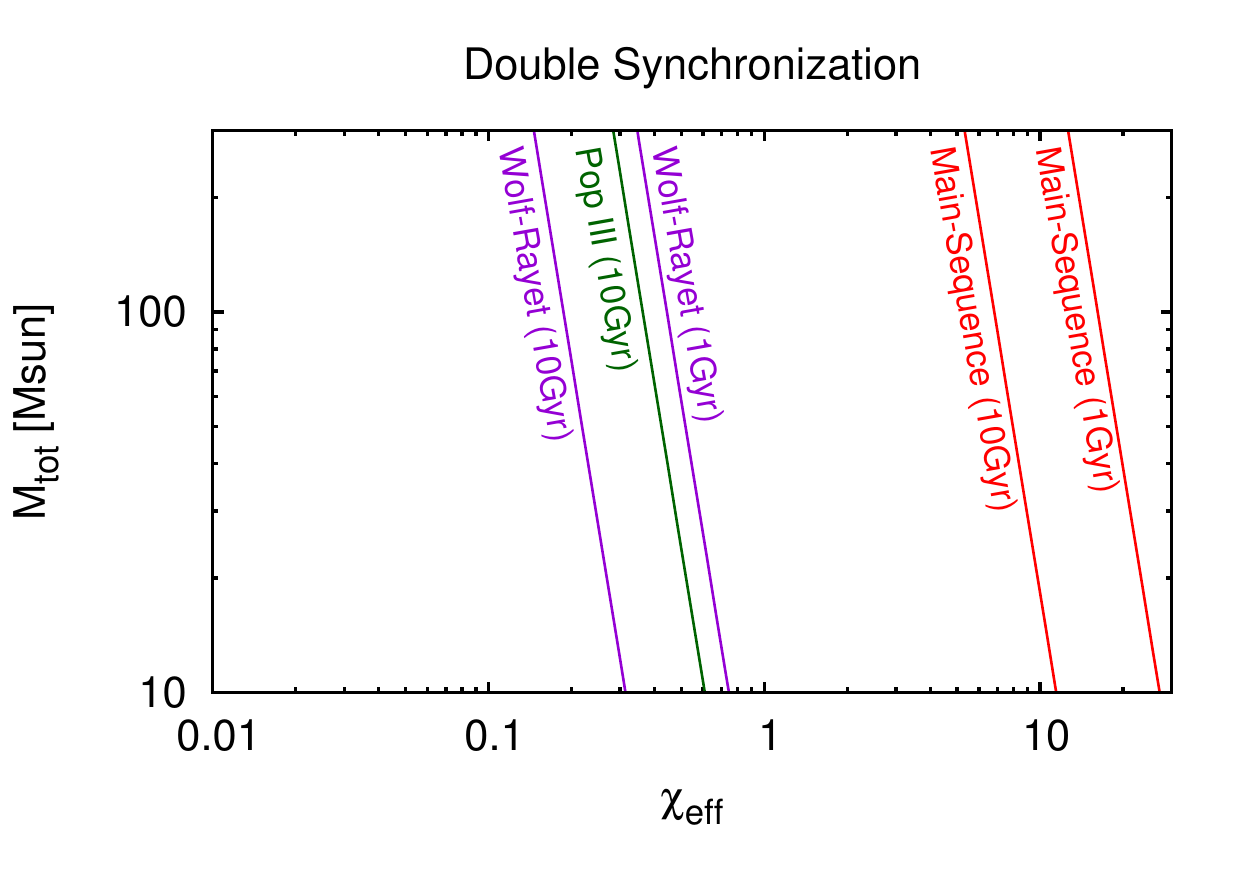}\\
         \includegraphics[width=70mm]{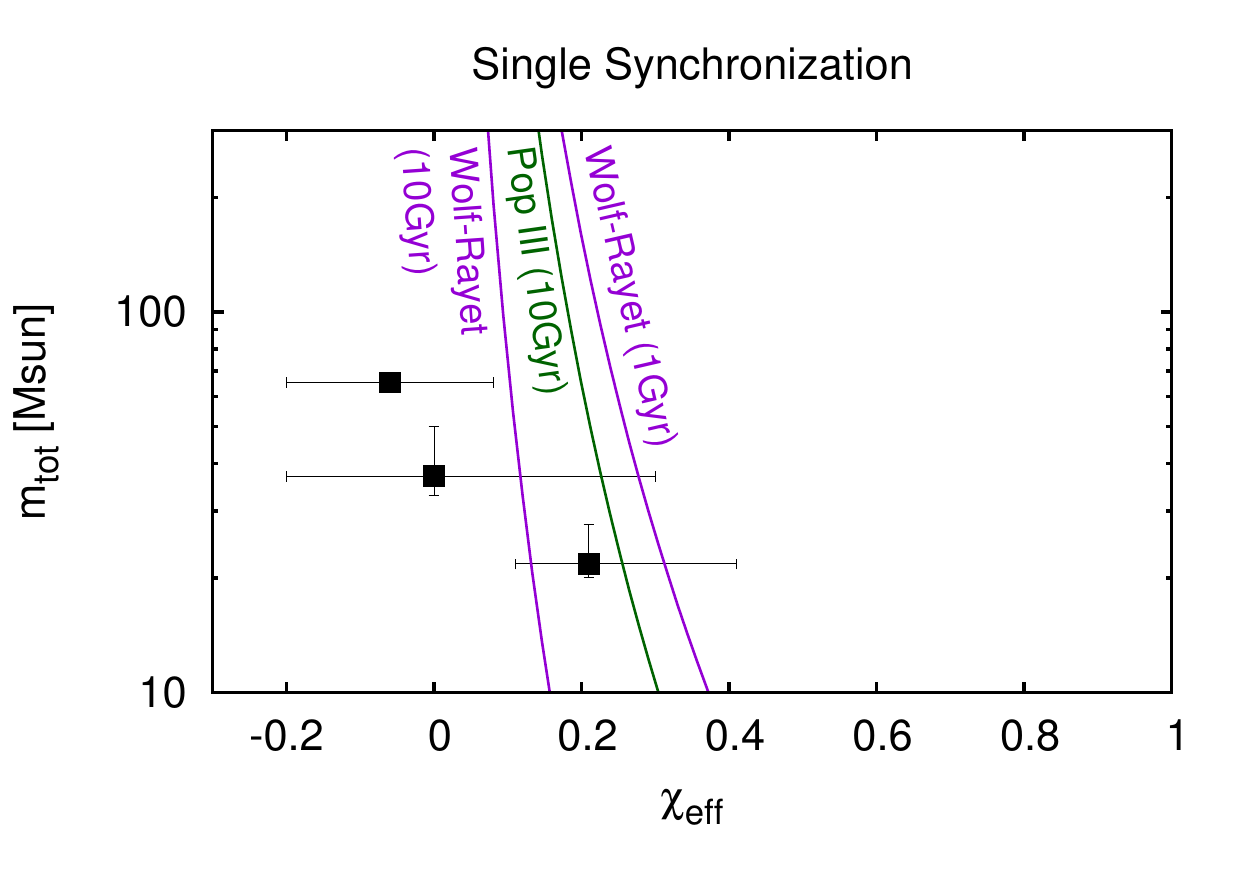}
     \includegraphics[width=70mm]{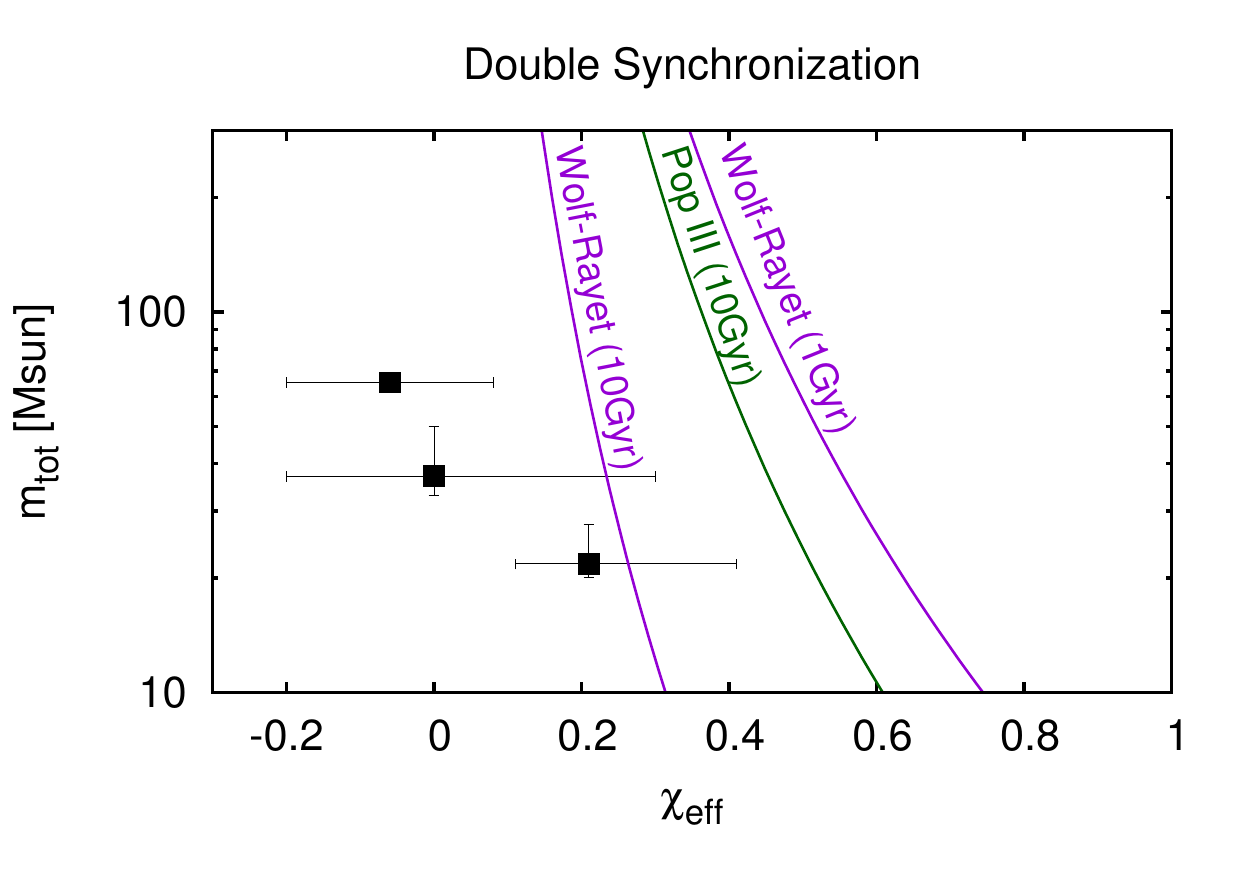}
    \caption{The spin and total mass of binaries in which  the stellar rotation is
    synchronized with the orbital motion. Here we consider different stellar
    models with coalescence times of $1$ or $10$~Gyr. The top panels show
    effective spin parameters in logarithmic scales. The bottom panel shows
    effective spin parameter ranging from $-0.3$ to $1$. Here we assume the the mass
    ratio of binaries is unity for the models. {When  comparing the models with the data in the bottom panels, we
    also assume that the BBH spin axes
    are aligned to the orbital axis.} The data are taken from
    \cite{abbott2016PhRvX}. 
}
    \label{fig:chi}
  \end{center}
\end{figure*}

Figure \ref{fig:MR} 
depicts the mass - radius relation of the different 
stellar models. Three diagonal lines depict the spin parameters
of these stars $\chi=0.1,\,1$, and $10$ (see Eq. \ref{stellarspin}),
if they are synchronized at the semi-major axis
for which a binary coalesces in a Hubble time. 
Also shown as a diagonal line is the critical  
stellar radius that exceeds the Roche limit
 in a binary system that  coalesces in  a Hubble time (see Eq. \ref{roche}).
Figure \ref{fig:chi} shows the relation between the
effective spin parameters of different stellar models with
the observed values from \cite{abbott2016PhRvX}.
Here we assume a mass ratio $q=1$,
the single (double) synchronization means that
one of (both) the  black holes in a BBH is formed from a synchronized star.
{When comparing the models with the measured values of $\chi_{\rm eff}$
we further assume that the spin axes of BBHs are aligned with 
the orbital axis (see \S \ref{sec:discussion} for caveats).}

For a given stellar model and a given coalescence time 
the spin parameter of the  synchronized progenitors
depends rather  weakly on the stellar mass. More specifically,
the spin parameter behaves as $\chi \propto m^{-0.225}$ 
for $R\propto m^{0.7}$, which is a typical dependence of
the radius of massive stars on the mass {\citep{tout1996MNRAS,kushnir2016MNRAS}.}
 Thus the  spin parameter   
reflects  the time delay  between the formation and the coalescence
irrespective of the BBH mass.

{\it (i) Main-sequence stars:} While we don't  expect a main-sequence 
star to collapse directly  to a black hole, we  begin with
main-sequence binaries  and show that these are ruled out. 
Main-sequence stars with masses $\gtrsim 10M_{\odot}$
can exist in a binary system with $t_c=10$~Gyr without exceeding
its Roche limit. 

Massive 
main-sequence stars in close binaries with $t_{c}\lesssim 10$~Gyr are synchronized
on timescales much shorter than their lifetime  (see Eq.~\ref{eq:Zahn}, 
where we used  the stellar structure of main-sequence stars at
the median point of their lifetime; \citealt{tout1996MNRAS,hurley2000MNRAS}).
Thus, main-sequence stars are  tidally synchronized. 
In fact, Galactic O-star binaries with orbital periods $\lesssim 10$~days
are likely tidally synchronized \citep{ram2015A&A}.
The spin parameter of such main-sequence  stars always exceeds unity.
Therefore we can rule out the possibility that the BBHs detected in
 LIGO's O1 run have been formed directly from the collapse of main-sequence stars.

If the BBHs formed via binary evolution beginning with 
two main-sequence stars, then  in
order to reduce the spin parameter significantly  the progenitors must have experienced
either a significant mass loss carrying out most  of their spin angular momentum (more than $95\%$)
or a significant decrease in the semi-major axis during their evolution.
The former may occur due to a wind or to mass transfer during
the late phase and the latter may occur during a  common envelope phase.
The natural outcomes of these processes are WR stars,
which we discuss  later in this and the following sections. 
This conclusion seems to be consistent with  stellar and
binary evolution  models (e.g. \citealt{belczynski2016Nature}).

{\it (ii) Red supergiant stars} are late massive stars with
an extended hydrogen envelope, in which the convection is deeply developed. 
These stars are located around the Hayashi line in HR diagrams,
where the corresponding temperatures are around $3000\,$--$\,4000$~K.
 Red supergiants have high luminosities and cool effective temperatures,
implying that they have large radii of $100$ to $10^3 R_{\odot}$.
BBHs arising from such wide binaries
never merge within a Hubble time so that we can robustly 
exclude the scenario that red supergiants are
the progenitors of merging BBHs just prior to the core collapse.

{\it (iii) Blue-supergiant stars} are massive stars at their late phase 
with a hydrogen radiative envelope (see, e.g., 
\citealt{langer1994A&A,meynet2011BSRSL,hirschi2004A&A}). 
Their radii can be $10-30R_{\odot}$, corresponding to high effective temperatures,
and can be smaller than the Roche limit of a binary with
a coalescence time of $10$~Gyr.  The spin parameter of blue supergiants
is always  much larger than unity if they are synchronized.
Therefore, these stars are  unlikely progenitors of  LIGO's O1 events. 
However, the synchronization time is quite sensitive to the structure of the envelope
and  hence it is somewhat uncertain.  We will  address this issue in a separate work.

{\it (iv) WR stars} are late phase massive stars that have lost most of their 
hydrogen envelope (see, e.g., \citealt{langer1994A&A,meynet2003A&A,maynet2005A&A}).
{Importantly, a few WR--black hole binaries that 
are likely to evolve to merging BBHs have been observed in nearby galaxies (see \citealt{pretwich2007ApJ, silverman2008ApJ} for IC10 X-1,
\citealt{carpano2007A&A,crowther2010MNRAS} for NGC 300 X-1, 
\citealt{bulik2011ApJ} for the inferred BBH merger rate, 
\citealt{liu2013Nature} for M 101 ULX-1, and see also \citealt{esposito2015MNRAS} for
more candidates). } Because of the lack of the hydrogen envelope,
the stellar radius is small. It is related to the mass as 
$R\approx R_{\odot} (M/10M_{\odot})^{0.7}$ \citep{kushnir2016MNRAS}. 
The spin parameters of BBHs formed via synchronized WR stars  are shown in
Fig.~\ref{fig:chi}. 
For systems with $t_c\sim 10$~Gyr, the spin parameters
can be as small as $0.1$. These values are 
consistent with the measured effective spin parameters of the LIGO's O1 events.
However, based on Eq.~(\ref{eq:kush}), WR stars are so compact that WR stars 
in binaries with $t_c\gtrsim 1$~Gyr  are not tidally synchronized within their lifetime.  
We will discuss further the spin parameters of WR progenitors in \S \ref{sec:spin}.

\begin{figure*}
  \begin{center}
        \includegraphics[width=70mm]{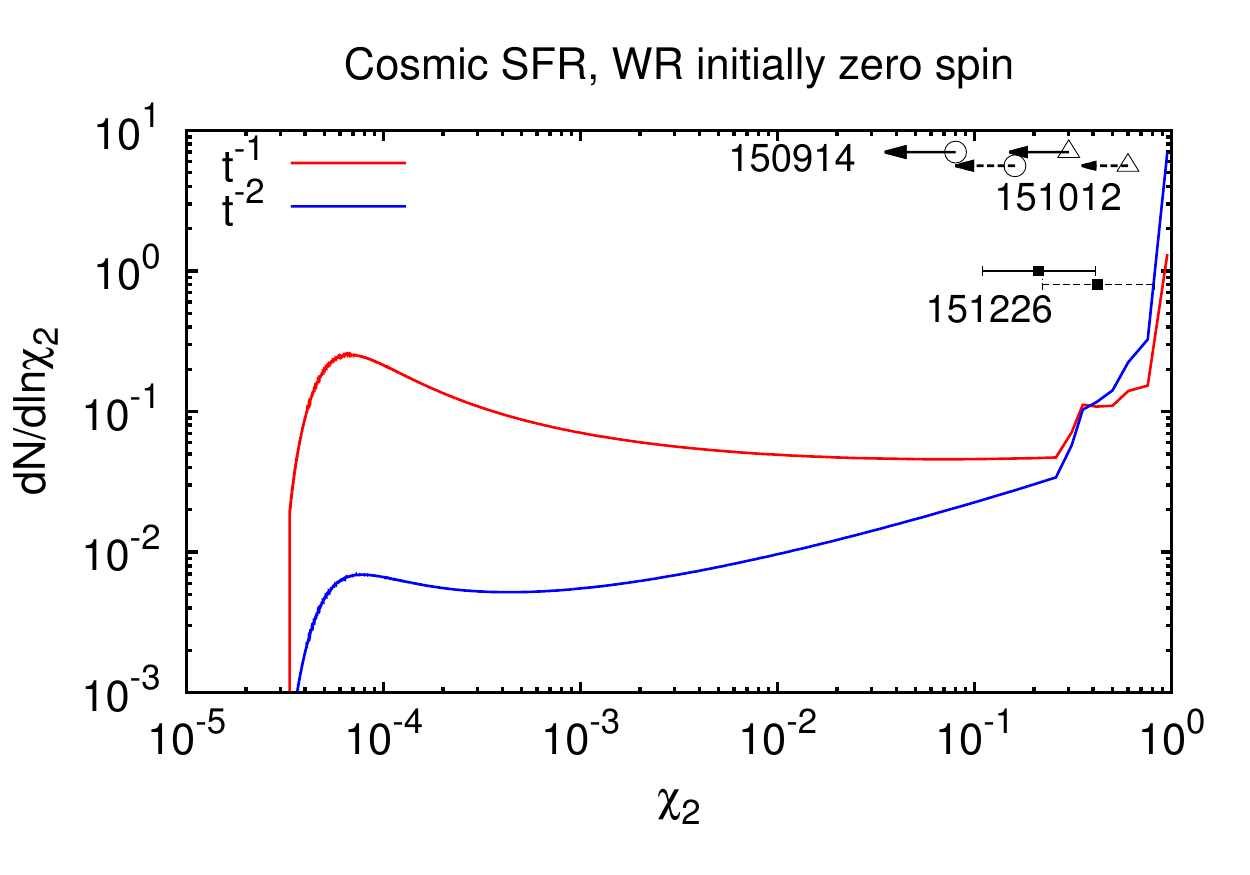}
        \includegraphics[width=70mm]{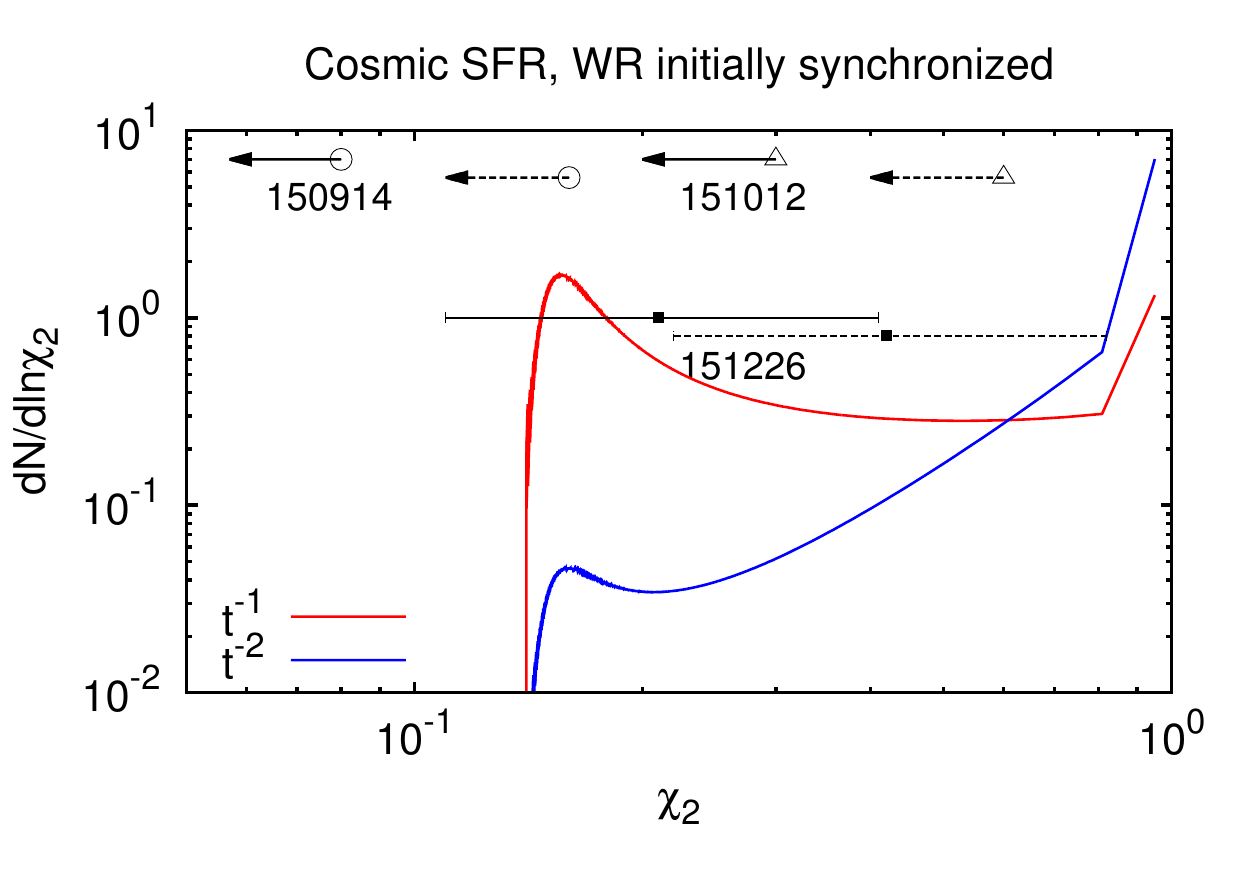}\\
         \includegraphics[width=70mm]{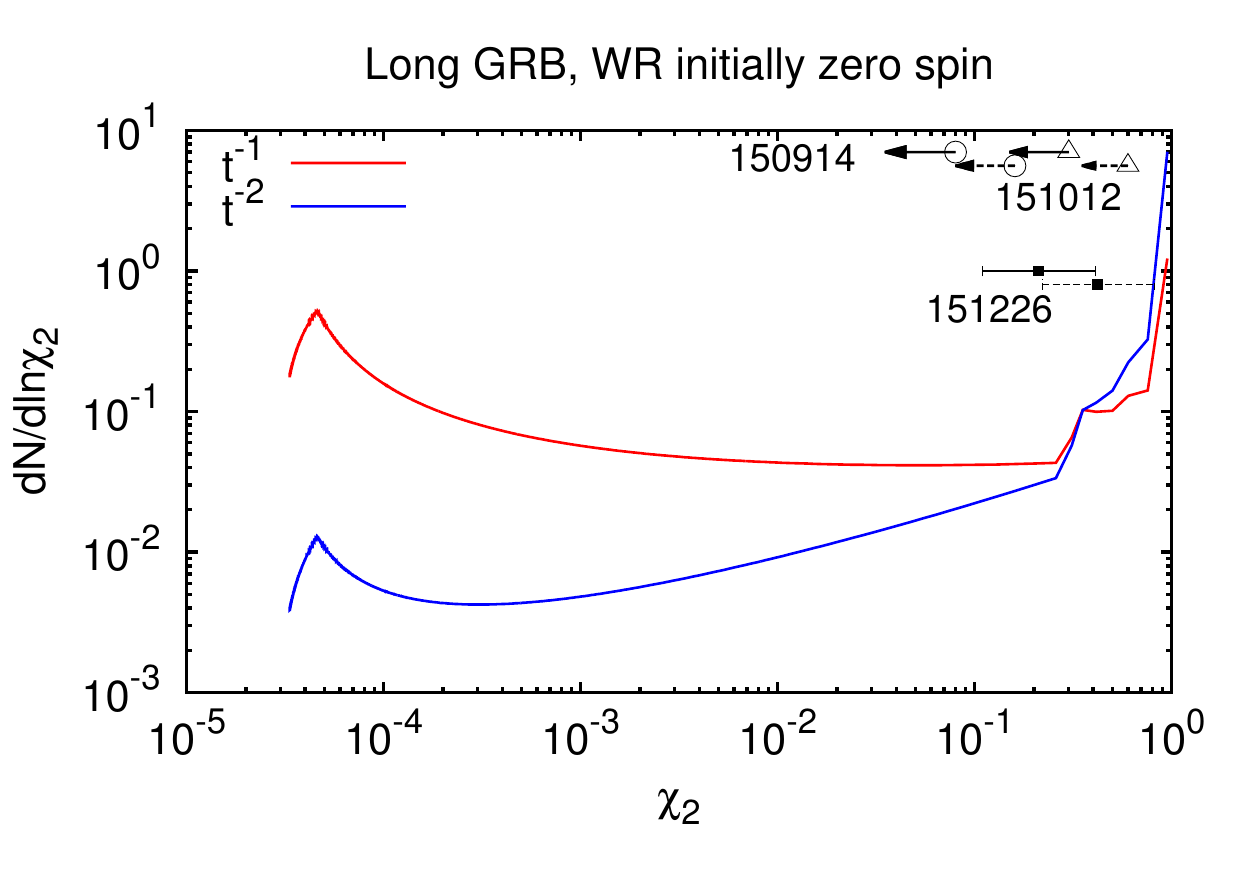}
        \includegraphics[width=70mm]{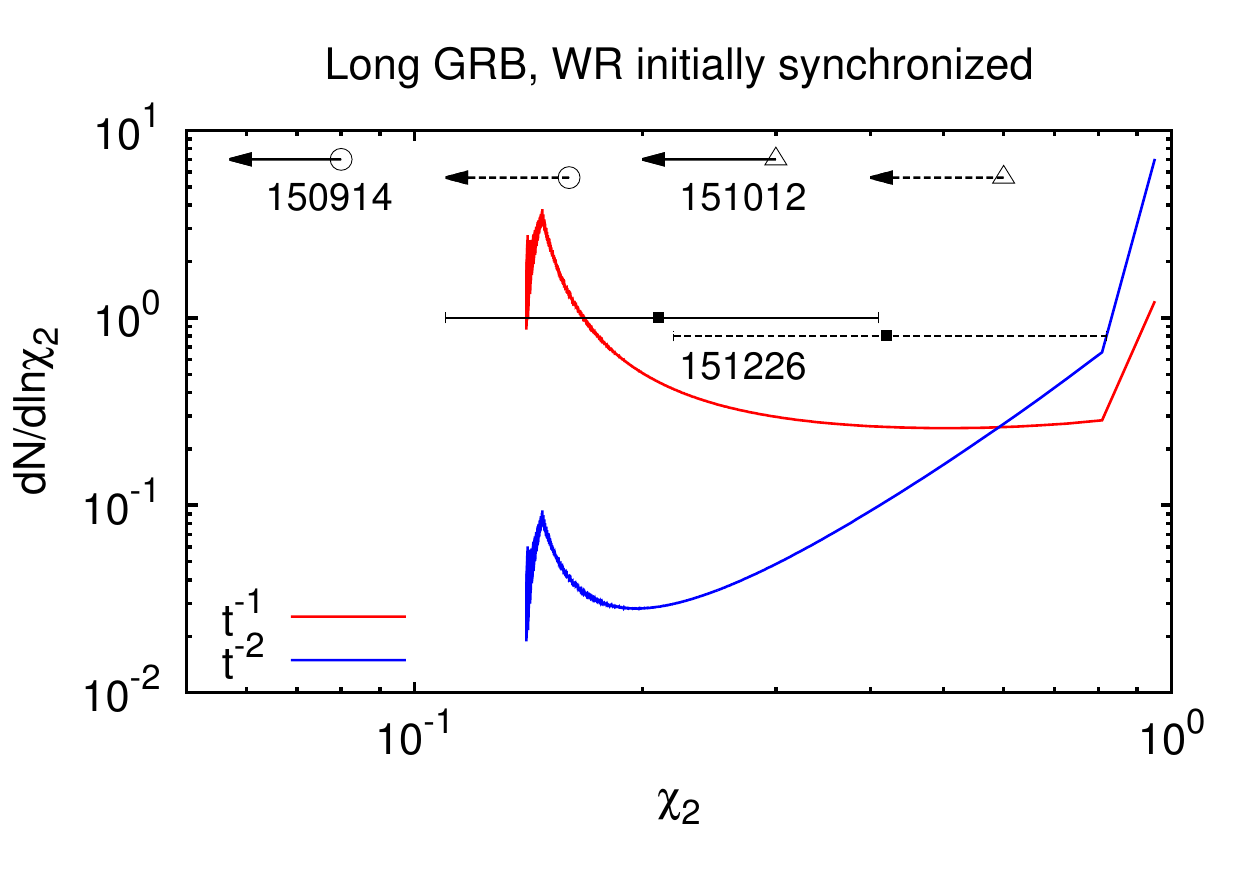}
    \caption{The spin distribution of BBH mergers at $z=0.1$ for
     BBH formation history that follows the cosmic star formation history (top panels)
    and the LGRB rate (bottom panels). 
   The distributions are calculated under the  assumptions that 
   the initial spin angular momentum of the WR stars vanishes, i.e., $x_{s,i}=0$ (left) and
   the WR stars are initially tidally synchronized, i.e., $x_{s,i}=1$ (right). We use 
   two different delay-time distribution $n=1$ and $2$ with a minimal time delay  of $10$~Myr. 
   We set the mass ratio, $q$,
   to be unity and the total mass to be $60M_{\odot}$. 
   The location of the peak at lower spins slightly shifts with changing the masses.
   Also shown 
   are the $\chi_2$ values inferred from the LIGO's O1 three detections.
   The measured effective spin parameters are translated to $\chi_2$
   assuming $\chi_1=\chi_2$ (solid line and arrows) and $\chi_1=0$. 
 {
 Here we assume that the BBH spin axes
    are aligned with the orbital axis  hence we 
    show the upper limits on the spin parameters of GW150914 and LVT151004. Note that the measured 
    spin parameters of these BBHs can be negative (see Table 1).}
}
    \label{fig:spin}
  \end{center}  
\end{figure*}

{\it (v) Population III stars} have  formed from pristine gas.
 They are typically massive stars with  twenty to a few hundreds  $M_{\odot}$  \citep{hosokawa2011Sci,hirano2014ApJ}. 
Their  radii are much smaller than those of  normal main-sequence stars because the core, that lacks metals, needs to be
compact in order to produce sufficient heats through
nuclear burning to support the stellar mass  (e.g. \citealt{omukai2003ApJ}).
Since
Pop III stars form only in the very early Universe around a redshift of $\sim 10$ (e.g. \citealt{desouza2011A&A}), 
BBH mergers at the local Universe that originate from Pop III stars have a  coalescence time of $\sim 10$~Gyr.
Using Pop III stellar
structure calculated by \cite{marigo2001A&A} we find that 
even though  Pop III stars are small, if they are synchronized, the spin parameters of BBH mergers in the local Universe
are between $0.2$ and $0.6$  (see  Fig.~\ref{fig:spin}).  
However,  the synchronization time of such systems is  
$\sim 10$ Myr, which is comparable to their lifetime,
so that Pop III stars in such binaries may not be fully synchronized.  
Therefore Pop III stars can be the progenitors of 
LIGO's O1 events.

The spin parameter of Pop III stars exceeds unity  if the synchronization 
occurs during the He-burning phase (see
 Fig.~\ref{fig:MR}).
However, these stars have a  convective core with a small radius 
and a shorter lifetime,
thereby  synchronization probably does not occur during
this Pop III He burning phase. Furthermore,
massive Pop III He stars exceed their Roche limit (see Fig. \ref{fig:MR}).
Therefore some fraction of the spin angular momentum may be removed
due to mass transfer in this phase.


\section{Long GRBs and BBH mergers}
\label{sec:LGRBs}
LGRBs arise from the core collapse of massive stars.
Supernovae associated with LGRBs are type Ibc, suggesting 
that the progenitors are stripped stars, e.g., WR stars. 
The progenitors' radii can be estimated from the properties of
the prompt emissions as follows. 
The plateau in $dN_{_{\rm GRB}}/dT_{90}$, where 
$N_{_{\rm GRB}}$ is the number of observed LGRBs and
$T_{90}$ is the
duration of prompt emission containing $90\%$ of its gamma-ray fluence,
indicates that the typical jet break-out time from the stellar surface is $\sim 15\,$s \citep{bromberg2012ApJ}.
This break-out time is related to the progenitor's parameters as \citep{bromberg2011ApJ}:
\begin{eqnarray}
t_b &\approx& 15~{\rm s}~\left(\frac{L_{j,{\rm iso}}}{10^{51}\,{\rm erg\,s^{-1}}}\right)^{-1/3}
\left(\frac{\theta_j}{10^{\circ}}\right)^{2/3} \nonumber \\ 
& &\times \left(\frac{R_*}{5R_{\odot}}\right)^{2/3}
\left(\frac{M_*}{15M_{\odot}}\right)^{1/3},
\end{eqnarray}
where $L_{j,{\rm iso}}$ is the isotropic jet luminosity, $\theta_j$ is the
jet's half opening angle,  $R_*$  and $M_*$ are the radius and mass 
of the progenitor. Note that these mass and radius that are  inferred from
the GRB observations of $L_{j,{\rm iso}}$, $\theta_j$, and $t_b$ are consistent
with the required properties of the progenitors of BBH mergers.

\begin{figure}
    \includegraphics[width=70mm]{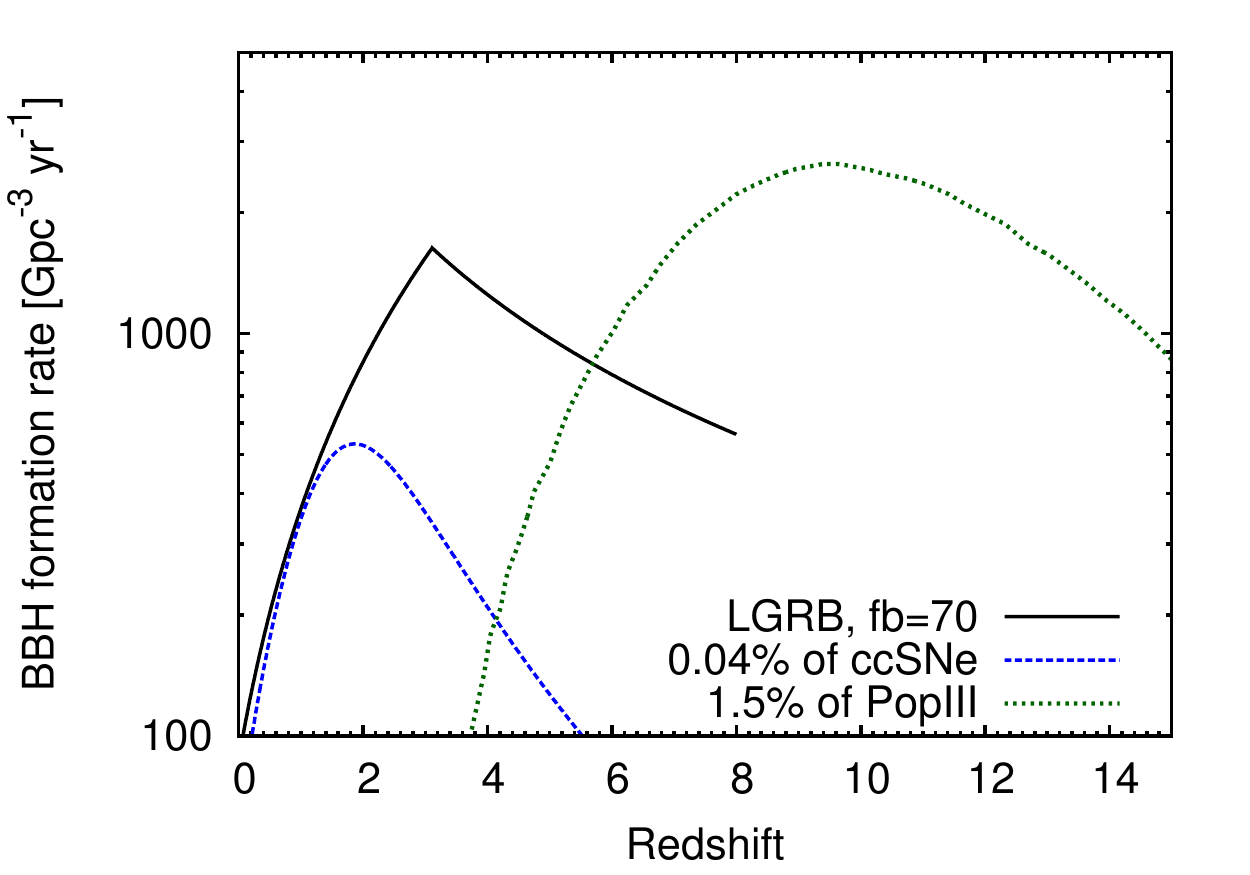}
    \caption{The  BBH formation rate under the assumptions that 
    it follows (i) the cosmic star formation history, (ii) the LGRB rate, and (iii) the Pop III star formation rate.
    The BBH formation rates is normalized to be $0.04\%$ of core collapse supernovae for (i; \citealt{madau2014ARA&A}),
    the LGRB rate with a beaming correction of $70$ for (ii; \citealt{wanderman2010MNRAS}), and 
    $1.5\%$ of Pop III star formation for (iii; \citealt{desouza2011A&A}).
    Here the mean stellar mass of Pop III stars is assumed to be $20M_{\odot}$.}
    \label{fig:bbh}
\end{figure}

The spin of the progenitor likely plays an essential role in the production of the 
GRB emission because the formation of a massive accretion torus
around a new-born black hole is required to produce the corresponding high luminosity 
jets (see, e.g., \citealt{macfadyen1999ApJ}).    The specific orbital angular momentum
at the inner most stable circular orbit is $j_{_{\rm ISCO}}=2\sqrt{3}$ for
Schwarzschild black holes and $2/\sqrt{3}$ for extreme Kerr black holes.
Here the angular momentum is normalized by the mass of the 
central black hole. The specific angular momentum of a mass element
of a rigidly rotating star at a radius $R$ on the equatorial plane is
\begin{eqnarray}
j(R) = \frac{\Omega R^2}{r_{g,_{\rm BH}}c} = \frac{\chi_*}{\epsilon}\left(\frac{R}{R_*}\right)^2\left(\frac{M_*}{M_{\rm BH}}\right),
\end{eqnarray}
where $r_{g,_{\rm BH}}$ is the gravitational radius of the
central black hole and $\chi_*$ is
the dimensionless spin parameter of the star.
The condition that the mass elements of the stellar core 
form an accretion torus is assumed to be $j(R_c)\geq j_{_{\rm ISCO}}$
or equivalently:
\begin{eqnarray}
\chi_*  & \gtrsim & 1.3 \left(\frac{\epsilon}{0.075} \right) \label{chi_c}
\left(\frac{R_c}{0.57R_{\odot}} \right)^{-2}\\ \nonumber
& &\times \left(\frac{R_*}{1.6R_{\odot}} \right)^{2}\left(\frac{M_{\rm BH}}{15M_{\odot}} \right)\left(\frac{M_*}{20M_{\odot}} \right)^{-1},
\end{eqnarray}
where we assume that the central black hole is a Schwarzschild black hole as 
a conservative choice 
and $R_c$ is the stellar core radius\footnote{{If we use as the condition of the disk formation
that the mass elements of the core surface have the  specific angular momentum of the marginally bound orbit,
which is $4$ for a Schwarzschild black hole as chosen in \cite{barkov2010MNRAS},
the critical  stellar spin parameter is larger by $15\%$ compared to Eq. (\ref{chi_c}).}}. 
The reference parameters are for a WR star 
taken from \cite{kushnir2016MNRAS}. Within this model,
LGRBs are produced by black holes only when 
the progenitor's spin parameter is larger than $\sim 1.3$,
and thus, the resulting black hole has a large spin (see also \citealt{barkov2010MNRAS}).
Using Eq. (\ref{spin}),
this condition can be translated to the coalescence time 
for a given stellar mass as $t_{c}\lesssim 0.2$ Gyr $(m/30M_{\odot})^{-13/8}$.
Therefore, if the delay-time  distribution is roughly $1/t$ and the minimal coalescence time is $\sim 10$~Myr,
one third  of BBH formation with $t_c<10$~Gyr have 
spins which may be large enough to produce LGRBs (see \S \ref{sec:spin} for the  delay-time  distribution). 
 Note that two LGRBs may lead to a single  BBH merger as  both the 
first and the second core collapses may produce GRBs, if
 they arise from a doubly synchronized system.

\section{The spin distribution and its redshift evolution of BBH mergers}\label{sec:spin}
We turn now to  the redshift-dependent
spin distribution of BBH mergers for different assumptions on the formation rate.
We focus on WR  progenitors. 
Here we assume that the spin parameter is $\chi_{{\rm BH}}={\rm min}(\chi_*,\,1)$.
We consider two different scenarios: (i) the WR stars are
synchronized at the beginning of the WR phase,  $x_{s,i}=1$; (ii) the initial spins of the WR stars are 
much smaller than the  synchronization spin, $x_{s,i} \approx 0$.
This latter initially low spin case,  $x_{s,i}=0$,
corresponds to an evolutionary path with a common envelope 
phase in which   
the semi-major axis shrikes significantly just prior to the beginning of
the WR phase. 
The spin distribution of BBH mergers can, therefore  be 
used to constrain whether or not a common envelope phase
plays an important role for the BBH progenitors. 

The BBH merger rate at a given redshift is given by a convolution of 
the cosmic BBH formation rate and the delay-time  distribution.
Here we assume a power law distribution of the delay time  {with a
minimal delay time}\footnote{The strong dependence of 
the merging time on the semi major axis suggest such a distribution with $n \lesssim 1$.}:
\begin{eqnarray}
\frac{dN}{dt_c} =\frac{N_0}{t_c^n}~~~~{({\rm for~t_c>t_{c,{\rm min}}})}, \label{delay}
\end{eqnarray}
 {where the normalization constant $N_0$ ensures that the integration of Eq.~(\ref{delay}) from
 $t_{c,{\rm min}}$ to a Hubble time is unity}
 and we consider  here
 $n=1$ or $2$ {and $t_{c,{\rm min}}=10$~Myr. Note that this kind of the  delay-time  distribution 
 is motivated by those of other astrophysical phenomena related to binary mergers.
 For instance, the delay-time  distribution of type Ia supernovae has $n\approx 1$ and $t_{c,{\rm min}}$  of 40~Myr
 to a few hundreds of Myr (\citealt{maoz2014ARA&A} and references therein)
 and that of short GRBs has $n\approx 1$ and $t_{c,{\rm min}}\approx 20$~Myr \citep{wanderman2010MNRAS,ghirlanda2016A&A}.}

 For the cosmic BBH formation rate,
 we consider two scenarios: (i) it is proportional to the cosmic star formation rate (SFR; \citealt{madau2014ARA&A}) and
 (ii) it equals the LGRBs \citep{wanderman2010MNRAS}. 
The normalization of the cosmic BBH formation 
of the LGRB scenario corresponds to the LGRB rate corrected
by a beaming factor of $f_b=70$. For the cosmic SFR scenario, our normalization corresponds to 
one BBH formed every $2.5\cdot10^{5}M_{\odot}$ stellar mass formation. This  roughly corresponds to a merging BBH formation rate that is $0.04\%$ 
of  the normal core-collapse supernova rate (assuming that one core collapse supernova occurs every $100M_{\odot}$ stellar mass formation   formation).
The cosmic BBH formation rates of these scenarios are shown in Fig.~\ref{fig:bbh}.


\subsection{The  spin distribution of BBH mergers in the local Universe}
Figure~\ref{fig:spin}  depicts the  spin distribution of merging BBHs at $z=0.1$ .  For simplicity,
we consider equal mass binaries. Also shown are 
the values and upper limits of the 
spin parameters $\chi_2$ inferred from  LIGO's O1 detections {assuming that the spin axes of BBHs are aligned with
the orbital axis. }
We consider  two cases relating the
effective spin parameters to the component spin $\chi_2$:
(i) a single synchronization: the primary\footnote{The primary black hole is the one formed at the first core collapse
for our definition. It is not necessarily  the
more massive one.} black hole's  spin is negligibly small, i.e.,  $\chi_2\approx 2\chi_{\rm eff}$, and
(ii) a double synchronization:
the primary black hole is also synchronized with a comparable spin parameter. In this case  $\chi_2\approx \chi_{\rm eff}$.

The spin distribution for  $n=1$ has two peaks. One is at
a high spin $\chi_2 \sim 1$ and the other is at a low spin $\chi_2 \sim 0.15$ 
($\ll 0.1$) for $x_{s,i}=1$ (for $x_{s,i}=0$).
The latter low spin peak corresponds to the spin parameter of BBHs
that are formed at the cosmic BBH formation peak.
The population is flat between the two peaks. This is simply because
of $dN/d\ln\chi_2 \propto dN/d\ln t_c={\rm const}$, inferred from Eq.~(\ref{spin}).
For  $n=2$, the population at higher spins ($\chi_2\gtrsim 0.3$)
dominates as expected from the fact that there are more BBHs 
with shorter coalescence times as $dN/d\ln\chi_2 \propto \chi_2^{8/3}$ 
for $\chi_2\gtrsim 0.2$. This feature is irrespective of
the assumptions on the initial synchronization parameters and
the cosmic BBH formation history. It  suggests that
a steep delay-time distribution with $n\gtrsim 2$ is inconsistent with
the observed spin distribution. 
Clearly, given the different assumption, 
the  spin parameter distribution should be between the
single synchronization with $x_{s,i}=0$ and the double synchronization with $x_{s,i}=1$.
 
The bimodal spin distribution, that we find,  is 
qualitatively similar to one found by  \cite{zaldarriaga2017}.
However, the peak at the high spin in our calculation is 
lower than the one of \cite{zaldarriaga2017}. This is because we use a BBH formation history
that peaks at a redshift of $2$--$3$ so that the merger events 
at the local Universe are dominated by  a population with
longer coalescence times and therefore they have  smaller spins. 
 
 \subsection{The redshift evolution of high/low spin BBH mergers}
\begin{figure*}
  \begin{center}
    \includegraphics[width=70mm]{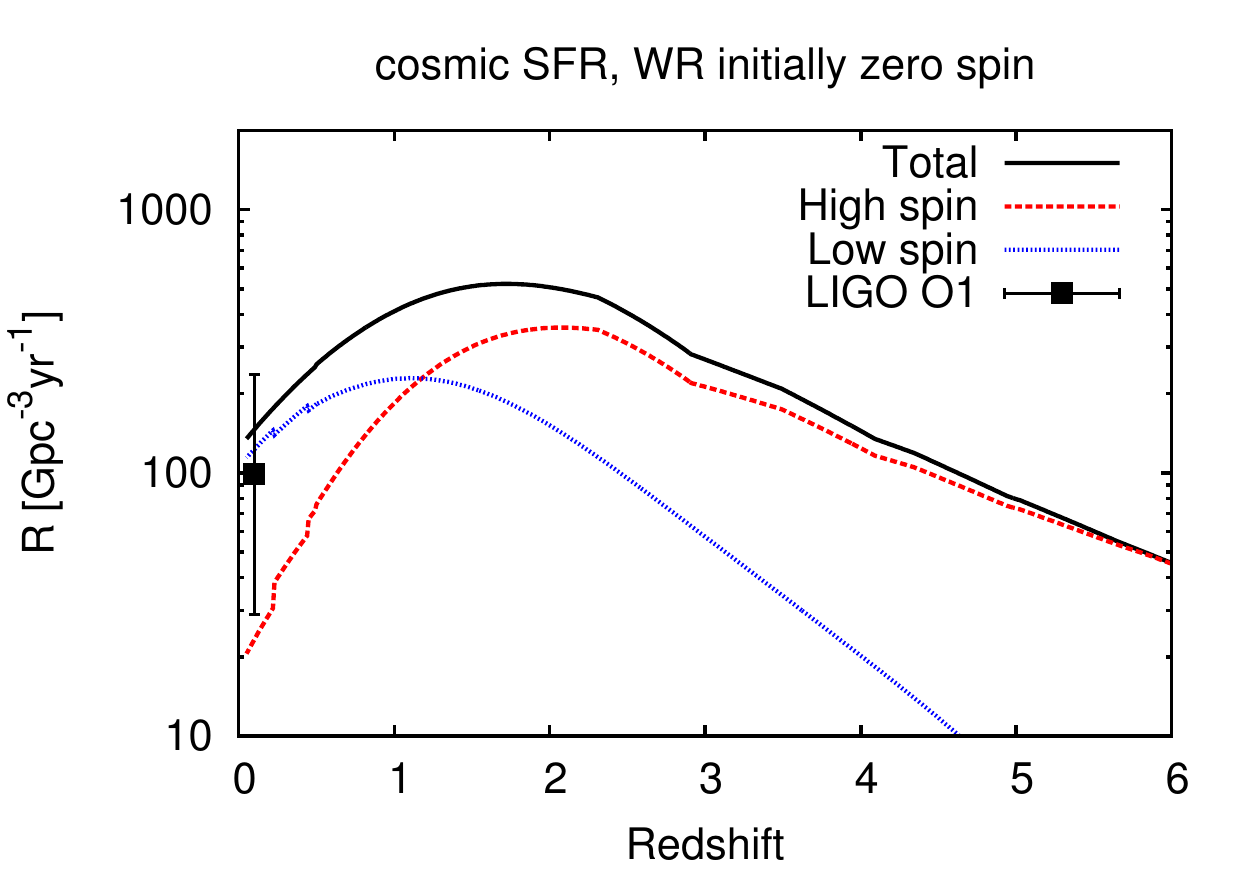}
        \includegraphics[width=70mm]{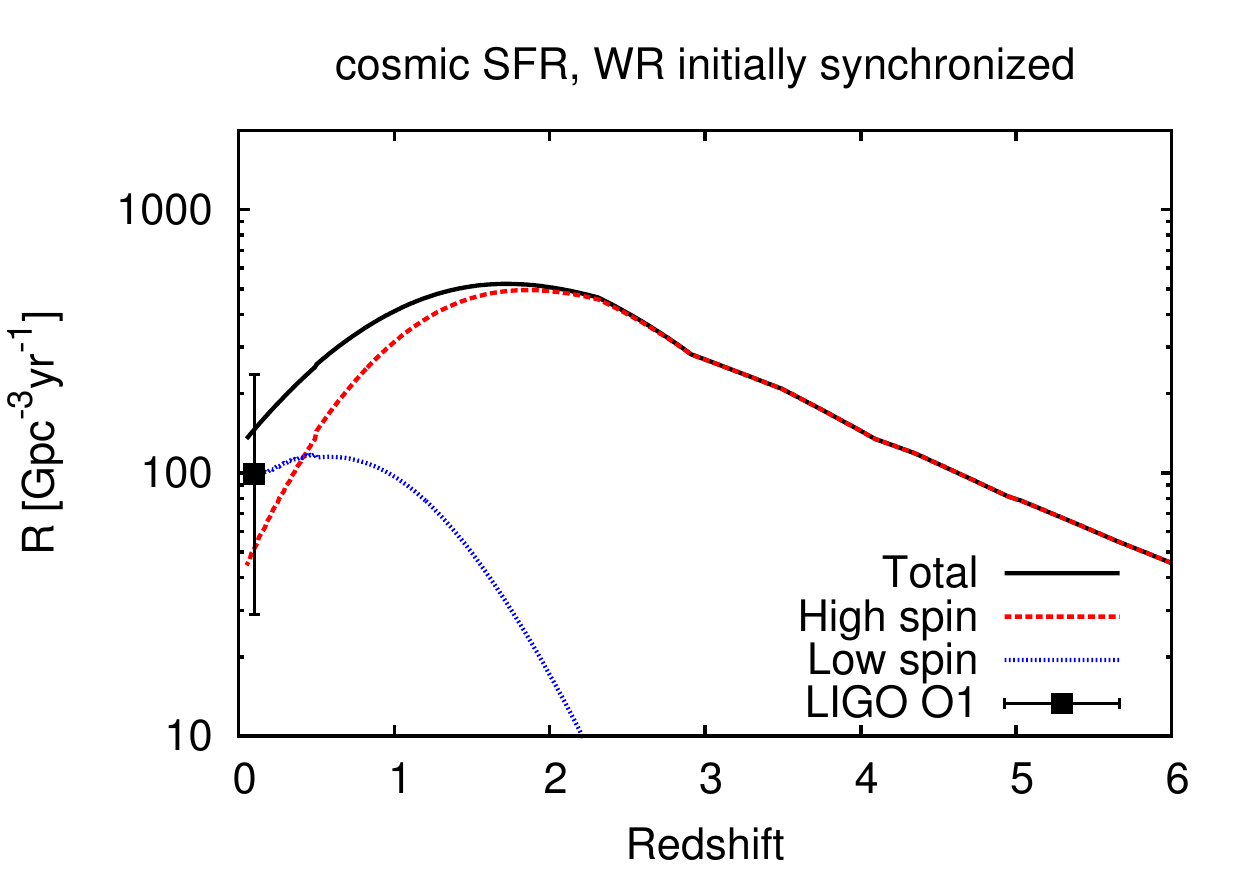}\\
        \includegraphics[width=70mm]{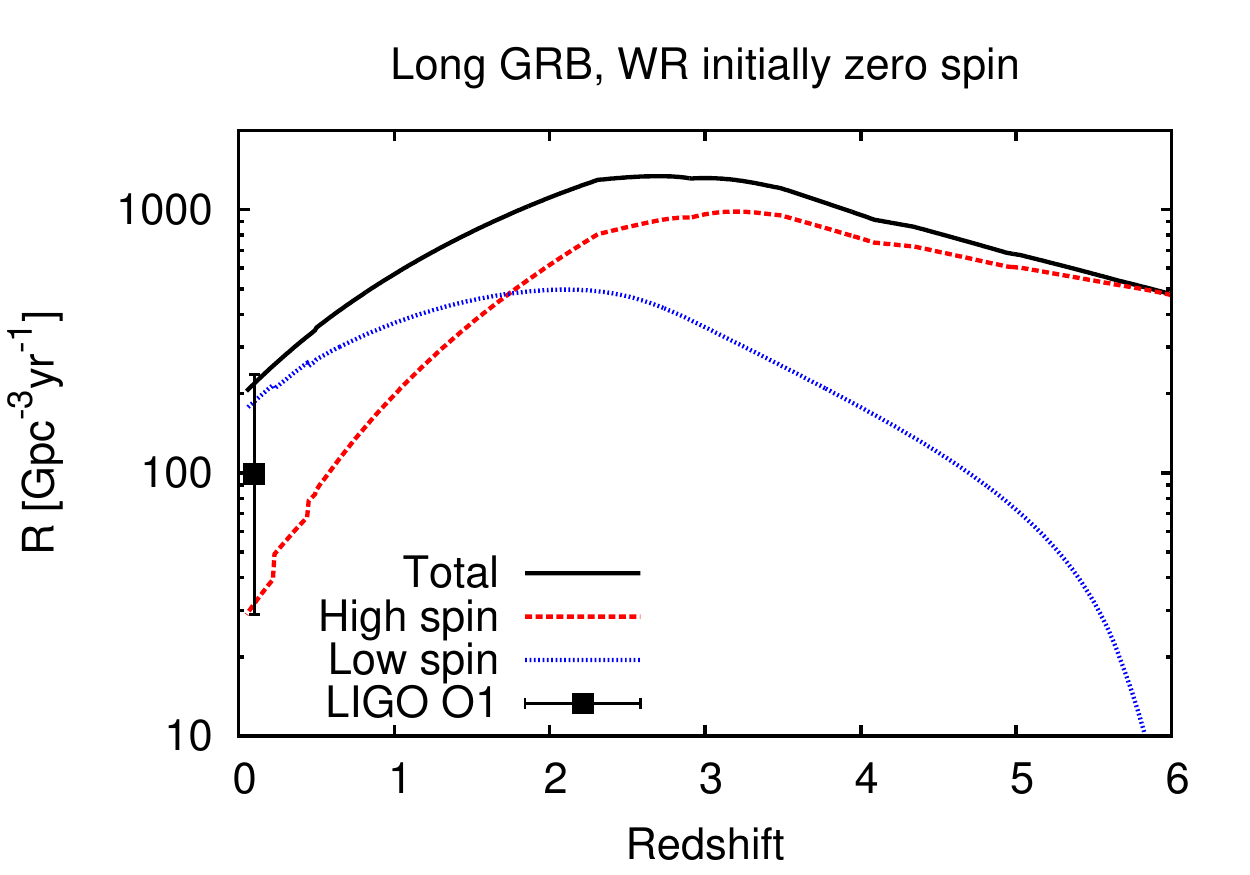}
        \includegraphics[width=70mm]{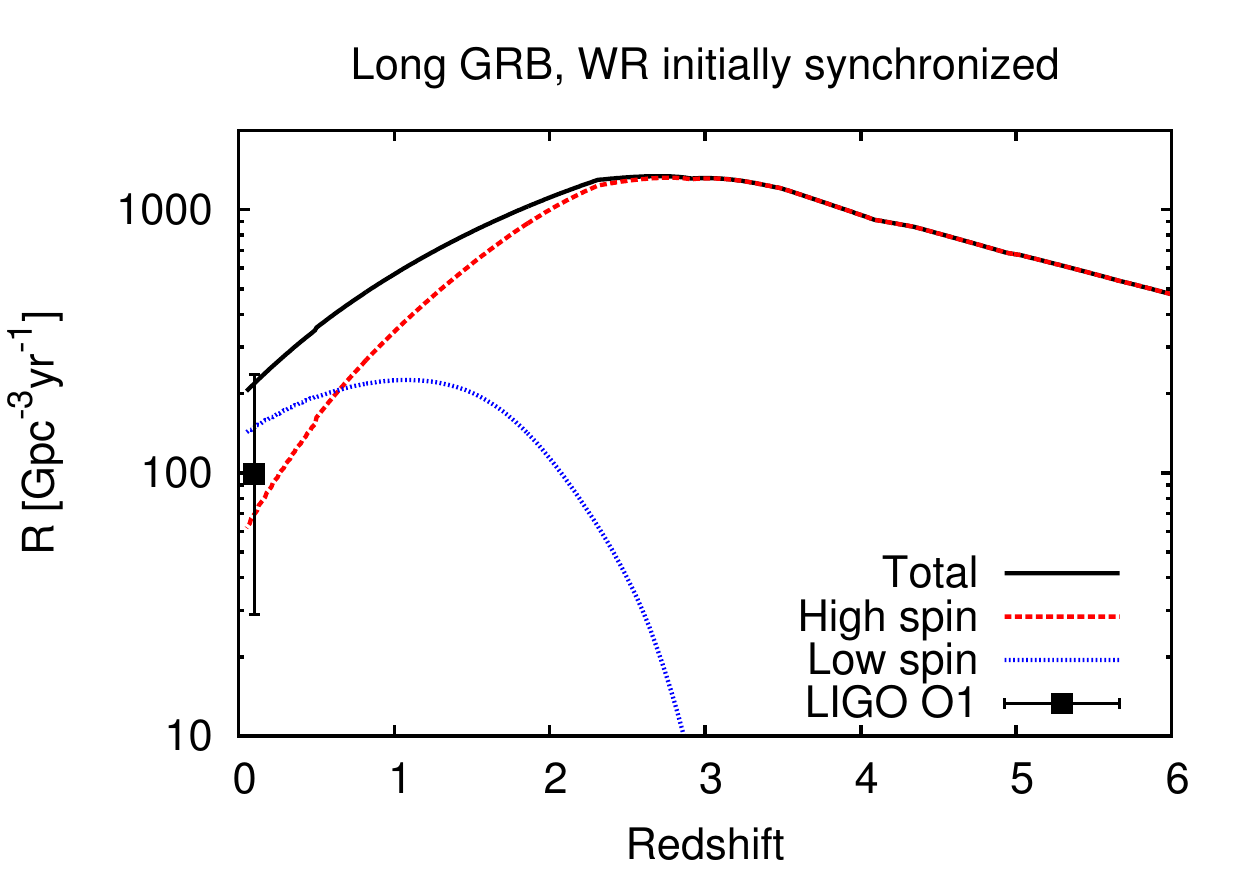}
     \caption{The redshift evolution of BBH mergers for the cases that
    the BBH formation follows the cosmic star formation history (top panels) and
    the LGRB  rate (bottom panels). We separate the mergers
    into the high  and low spin populations with a threshold spin of $\chi_2=0.3$.
    Here we assume a delay-time  distribution with $n=1$ and
    a minimal time delay  of $10$~Myr. The total merger rate in the local Universe 
    estimated by \cite{abbott2016PhRvX} is shown as a square.
}
    \label{fig:redshift}
  \end{center}
\end{figure*}

Figure~\ref{fig:redshift}  shows the redshift evolution of 
the BBH merger rate for  the cosmic SFR and  the LGRB scenarios.
We divide the BBH mergers into two classes (i) high spin ($\chi_2>0.3$)
and (ii) low spin ($\chi_2<0.3$). This threshold spin value  corresponds to
coalescence times of $0.3$~Gyr and $1.5$~Gyr for $x_{s,i}=0$ and $x_{s,i}=1$,
respectively. Because of the longer delay of the lower spin population 
the high spin BBH mergers predominately occur at higher redshifts.
In all cases, the merger rate of the low spin population is larger than 
that of the high spin one in the local Universe.
An interesting feature is that the merger rate of the 
high spin population starts to dominate over the low spin one at
a redshift of $\sim 0.5$--$1.5$.
 Mergers at such redshifts 
could be detected by upgraded GW detectors in future.

The  BBH merger history, which is a  convolution of the formation history with
a delay-time  distribution,
 is not very sensitive
to the assumption of the BBH formation history,
i.e., LGRB or cosmic star formation history,
as long as the latter peaks  around a redshift of $2\,$--$\,3$.

To produce LGRBs, the 
black holes should have  extreme spins (see  \S \ref{sec:LGRBs}). This is not the case for LIGO's O1 detections\footnote{
{Unless the spin-orbit misalignments are significant. Note also that the observed $\chi_{\rm eff}$ of GW151226 does not rule out the 
possibility that the less massive black hole has a spin parameter of order unity.}}.
However,  as noted earlier,   BBHs with high spins have short merger times   thereby
we do not observe most of these mergers in the local Universe. 
If the delay-time  distribution is $\sim 1/t$
with $t_{\rm min}=10$~Myr, 
we expect that the current  event rate of BBH mergers with extreme spins
is $\le 20\%$ of the total merger rate. The corresponding high spin  BBH merger
rate is  $\lesssim 20^{+28}_{-14}~{\rm Gpc^{-3}\,yr^{-1}}$.
On the other hand,  the local rate of LGRBs is $\sim 91^{+42}_{-49}~{\rm Gpr^{-3}\,yr^{-1}} (f_b/70)$,
where $f_b$ is a beaming correction factor \citep{wanderman2010MNRAS}.
Note that the LGRB rate should be compared with twice of the high spin BBH merger rate
for the double synchronization case.
These rates are consistent with each other within the admittedly large uncertainties.
This suggests that it is possible that the two phenomena share same progenitors.

\subsection{Pop III BBH mergers}
\begin{figure}
  \begin{center}
    \includegraphics[width=70mm]{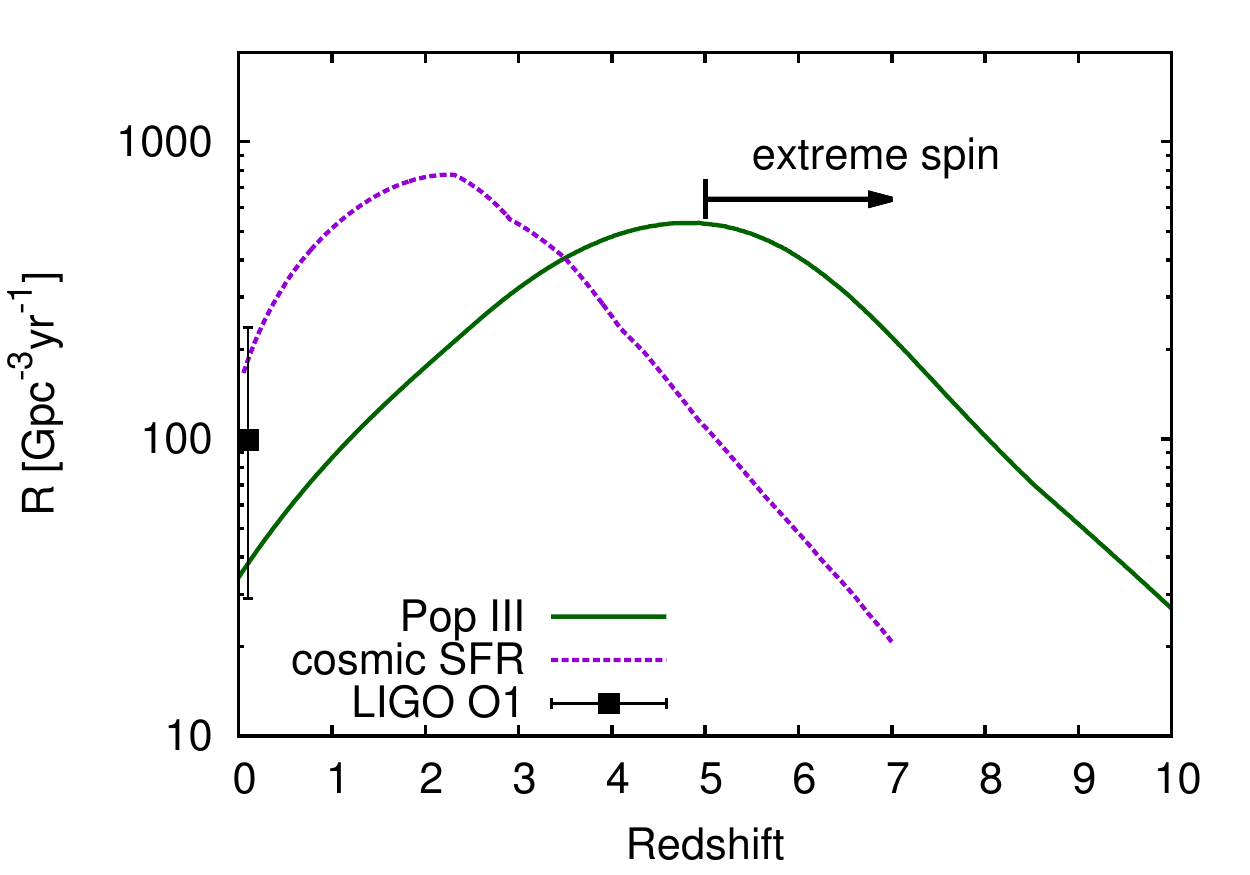}
    \caption{The same as Fig.~\ref{fig:redshift} but for the Pop III scenario.
    Also shown is the redshift evolution of the cosmic SFR scenario for a comparison.
    An arrow depicts the redshift where BBH mergers with extreme spins start to dominate
    the event rate. 
}
    \label{fig:pop3}
  \end{center}
\end{figure}

Figure \ref{fig:pop3} shows the
redshift evolution of BBH mergers for the Pop III scenario. 
Here we use a Pop III star formation rate derived by 
\cite{desouza2011A&A}\footnote{This Pop III star formation rate
seems the maximum allowed by the Planck observations of the electron scattering opacity to 
the cosmic microwave background within  two $\sigma$  (see, e.g., \citealt{visbal2015MNRAS} for details).}.
 We normalize the Pop III BBH formation rate such that
$1.5\%$ of Population III stars form BBHs with  coalescence times
less than a Hubble time \citep{inayoshi2017}. Here we assume a mean stellar mass of $20M_\odot$,
 a  delay-time  distribution with  $n=1$ and a minimal 
time delay  of $0.4$~Gyr. 
This minimal time delay  roughly corresponds to
the minimal semi-major axis for which  the radius 
of Pop III main-sequence stars is smaller than
the Roche limit.

The redshift evolution of Pop III
BBH mergers is significantly different from other astrophysical scenarios.
It increases up to $z\sim 5$, which is 
beyond the peaks of the cosmic star formation history 
and the  LGRB rate. This by itself can be used to distinguish
this scenario from  the others (see also \citealt{nakamura2016PTEP}). 
Another prediction of this scenario is that the spin parameters
of BBH mergers at higher redshifts above $\sim 4\,$--$\,5$
may be dominated by an extreme spin population with $\chi_{2}\sim 1$.
Clearly, significant improvements in GW detectors is needed to detect such events.

\section{Caveats}\label{sec:discussion}

{\it Uncertainties in the synchronization:}
The tidal synchronization relevant to 
the BBH progenitors is due to  dynamical tides that are excited above the convective core
and dissipate in the radiative envelope \citep{zahn1975A&A,goldreich1989ApJ,kushnir2017MNRAS}. 
Once a stellar structure is given, one can calculate the tidal torque on the star.
However,   massive stellar envelopes might be turbulent and unstable.
In such cases, synchronization due to an equilibrium tide in the envelope can be
more efficient (see, e.g., \citealt{toledano2007A&A,detmers2008A&A}). In this case 
the synchronization time  behaves as  $\propto q^{-2}(a/R)^6$.
This additional effect will speed up the synchronization. {Note also that
we have assumed circular orbits in this paper. One should use estimates of
 the synchronization  time including resonant excitations of g-modes when considering elliptic orbits \citep{witte1999A&A}.  }
But these effects are beyond the scope of this paper. We will address this issue in a separate work.

The  angular momentum loss due to winds  is  uncertain and it depends on the stellar metallicities.
While our results depend on the wind strength,  the qualitative 
results in this paper are robust. Indeed, \cite{zaldarriaga2017}
show the robustness of this spin argument for WR progenitors 
for different wind parameters.

{\it Mass loss and natal kick at the core collapse:}
We have assumed here that the mass of black holes is
identical to that of the collapsing stars. This assumption 
is likely valid as long as the spin parameter of the progenitors does not
exceed unity \citep{sekiguchi2011ApJ,oconnor2011ApJ}. When the progenitor's spin exceeds unity,
a fraction  of the progenitor's mass is ejected carrying the excess angular momentum, and hence,   the 
black hole has a mass smaller than the progenitor's mass \citep{barkov2010MNRAS}.
For WR stars, this effect is expected to be small since
their maximal spin parameter does not significantly exceed unity.

One of the major concerns about the spin argument is that 
it is assumed that the direction of the spin angular 
momentum is parallel to   the orbital angular momentum.
This assumption is crucial because  GW measurements are  insensitive to
the spin parameters perpendicular to the orbital angular momentum. The tidal torque
always works toward the orientation of the stellar spins to be parallel
to the orbital angular momentum. Other effects
of the binary interaction, e.g., mass transfer, also change the spin
components parallel to the orbital angular momentum.
{It  has been suggested  that the progenitor receives a natal kick  
during the core collapse (e.g. \citealt{janka2013MNRAS}).  This  may cause a misalignment
between the spin axes and the orbital axis. However, to significantly change the direction of the orbital angular momentum one
needs a kick $\gtrsim 500$ km/s. While such kicks have been observed at the high end of pulsar kicks 
 we do not expect such large 
natal kicks during  black holes formation. 
The fraction of the ejected mass  to the remnant mass in this case  is expected to be 
much smaller for black hole formation than the ratio in neutron star formation (see \citealt{janka2013MNRAS} and
also \citealt{rodriguez2016ApJ} for a detailed study of the  distributions $\chi_{\rm eff}$ including the natal kicks). }
Furthermore, observations of low mass X-ray binaries 
show no evidence of strong natal kicks of black holes (see, e.g., \citealt{mandel2016MNRASb}).
These  suggest typical values  that are  much smaller than the orbital velocities of the BBH progenitors.
Therefore, we expect that BBH natal kicks may not affect significantly the spin
components parallel to the orbital axis of the black holes 
and hence the results of our analysis (see also discussions in \citealt{abbott2016ApJ}).

{\it Mass transfer and  Common envelope phases:}
We considered  two scenarios, (i)  single synchronization and
(ii)  double synchronization. The spin of the black hole formed at
the second core collapse of a binary is conserved as long as there
is no significant mass accretion from the interstellar medium.
Therefore, the spin parameters in the single synchronization
case may be quite robust. On the contrary, the spin of the black hole
formed at the first core collapse can change from the value at
the birth of the black hole due to the mass accretion from the companion.
Moreover, the semi-major axis may further change after
the first core collapse due to a common envelope phase.
If this occurs, the spin parameter of this black hole
has nothing to do with the initial semi-major axis of BBHs.
Thus, the double synchronization case involves some
uncertainties, or equivalently, the spin parameter of
one of the black holes in BBH mergers is not
well constrained by the tidal synchronization argument.

{\it Spin reduction due to the Blandford-Znajek process:}
{One of the possible mechanisms powering GRB central engines
is the Blanford-Znajek process, in which the rotational energy of a central
black hole is removed through magnetic fields 
and an ultra-relativistic jet is launched with this energy \citep{brandford1977MNRAS}. 
While we still don't know whether or not this process works in
collapsing massive stars and what the back reaction of this process 
on the central black hole is,
if this process removes a significant amount of the rotational energy,
the spin of the black hole is reduced. }


\section{Conclusions}\label{sec:conclusion}

{The spins (projected  on the orbital angular momentum axis)  of the merging black holes observed by 
LIGO O1 run are rather small. We have examined the implications of these observations on the progenitor scenarios  
in which BBHs arise from isolated field binaries.  Our analysis was done under the assumption that this projection  indeed reflects the final spin of the progenitor star, just before it collapsed to a black hole.  As the expected mass ejected during the formation of the black hole is rather small compared with the remnant black hole we do not expect  a strong natal kick and hence a significant change in the black hole's spin or in the orbital angular momentum (see \citealt{rodriguez2016ApJ} for the effects of natal kicks on the
effective spin parameters).}

We have studied the spin distribution and its redshift evolution 
based on the tidal synchronization argument \citep{kushnir2016MNRAS}. 
We find that  massive main-sequence progenitors, whose  
semi-major axis is small enough to merge within a Hubble time,
the tidal synchronization occurs
on timescales much shorter than their lifetime.
As a result, the spin parameters of such main-sequence stars
exceed unity. Given the fact that the {aligned} spin parameters of
the three LIGO's O1 events measured via the GW signals are significantly 
less than unity, we can rule out the possibility that these BBHs 
 are formed directly from the collapse of main-sequence stars.
This also indicates that,
if the BBHs formed via binary evolution beginning with 
two main-sequence stars, the progenitor binary systems must experience
either a significant loss of their spin angular momentum (more than $95\%$)
or a significant decrease in the semi-major axis during their evolution.
This conclusion  is consistent with current  stellar and
binary evolution studies (see, e.g., \citealt{belczynski2016Nature} ).

Among known stellar objects,  WR stars seem to be  the only  possible progenitors
of the BBH mergers.  
We consider the spin distribution and redshift evolution
of BBH mergers formed via WR progenitors, taking the synchronization, mass loss,
and stellar lifetime, into account. Here we assume  that
the cosmic BBH formation history is proportional to either 
the cosmic SFR or to LGRB rate (LGRBs are also formed from WR stars)
with two different delay-time  distributions.
We find that a steep delay-time distribution $\propto 1/t^2$ 
predicts too many  BBH mergers with extreme spins $\chi \sim 1$.
This is inconsistent with the LIGO's O1 events. On the contrary,
for the delay-time  distribution of $\propto 1/t$, the rate of 
BBH mergers with  low spins ($\chi \lesssim 0.3$) dominates over the one with high spins ($\chi \gtrsim 0.3$)
in the local Universe. The ratio of the high spin mergers to
the low ones increases with the cosmological redshift and the
high spin population begins to dominate at a redshift of $0.5\,$--$\,1.5$.
This feature may be observable by GW detectors
network in future. 

The BBH merger rate density inferred from LIGO's O1 run is compatible 
with that of LGRBs. Motivated by this, we considered the possibility that
BBH mergers and LGRBs share the same progenitors, i.e,
LGRBs are produced at the core collapse of stars in a close binary 
which eventually evolves to a BBH with a coalescence time of  less than a Hubble time.
We show that a stellar spin 
parameter of $\gtrsim 1.3$, or equivalently a coalescence time of $\lesssim 0.2$ Gyr, 
is required  for WR progenitors  in order that
a fraction of the stellar core forms an accretion disk around the central black hole.
Assuming a delay-time  distribution of $1/t$ with the minimal delay-time   of $10$~Myr, 
we expect that  the LGRB rate is about one third of the BBH formation rate. 
Because  BBHs
with such extreme spins predominately merge at high redshifts,
it is still possible that BBH mergers and LGRBs share the same
progenitors even though the {aligned} spin parameters of the LIGO's O1 events
are significantly less than unity. We extrapolate the total BBH merger rate
with low spins inferred from LIGO's O1 run to the extreme spin population
based on the WR progenitor scenario and show that the BBH merger rate with
extreme spins is $20\%$ of the total rate or less. 
This can be tested in the near future with further observations of BBH mergers.

We also considered the hypothetical Pop III BBH merger scenario. 
As these BBHs formed at  high redshifts around $z\sim 10$,  in the local Universe,
BBH mergers from Pop III stars   always have  a time delay 
of $\sim 10$~Gyr. If they arise from synchronized stars, this corresponds to  the BBH
 spin parameters of $0.2\,$--$\,0.6$.
 However, it is not clear that these Pop III binaries are fully synchronized during 
 their main-sequence phase as the synchronization time
 is comparable to their lifetime.
 Furthermore, a fraction of the spin angular momentum may be
removed during the stable  mass transfer in the late phases and this may reduce
the spin parameters (see \citealt{inayoshi2017}). Therefore we conclude that the Pop III star scenario can be
consistent with the low {aligned} spins of the three LIGO's O1 events. 
In this scenario the BBH merger rate increases with the redshift up to $z\sim 5$
and we expect BBH merges with extreme spins beyond a redshift of $4$.
These are unique observable features of this scenario.

To summarize, we have shown  that the  low {aligned} spins of the BBH mergers observed in LIGO's O1 run are consistent with WR progenitors. Those are also  progenitors of LGRBs and given the comparable observed rate it might be that LGRBs arise when the WR progenitors collapse
to form the observed BBHs. While the observed spins are slightly lower than expected, Pop III stars cannot be ruled out either. 
Both scenarios predict that some high {aligned} spin BBHs should be discovered as well. If these are not discovered within LIGO's coming runs,  then the 
observations will imply that it is unlikely that  LIGO's BBHs have been  formed via regular binary stellar evolution channels,
and then, the capture in dense environments (clusters
or galactic cores) or primordial origin will be preferred.

\acknowledgments
We thank Maxim Barkov, Matteo Cantiello, Sivan Ginzburg, James Guillochon, Kohei Inayoshi, 
Tomoya Kinugawa, Itai Linial,  Masaru Shibata, Maurice van Putten, and Roni Waldman for useful 
discussions and comments. KH is supported by the Flatiron Fellowship at
the Simons Foundation.
The research was supported by an advanced 
ERC grant TReX and by the ISF-CHE I-Core center of excellence for research in Astrophysics. 

\newcommand{\jcap}{JCAP}

\begin{thebibliography}{74}
\expandafter\ifx\csname natexlab\endcsname\relax\def\natexlab#1{#1}\fi

\bibitem[{{Abbott} {et~al.}(2016{\natexlab{a}}){Abbott}, {Abbott}, {Abbott},
  {Abernathy}, {Acernese}, {Ackley}, {Adams}, {Adams}, {Addesso}, {Adhikari},
  \& et~al.}]{abbott2016ApJ}
{Abbott}, B.~P., {et~al.} 2016{\natexlab{a}}, \apjl, 818, L22

\bibitem[{{Abbott} {et~al.}(2016{\natexlab{b}}){Abbott}, {Abbott}, {Abbott},
  {Abernathy}, {Acernese}, {Ackley}, {Adams}, {Adams}, {Addesso}, {Adhikari},
  \& et~al.}]{abbott2016PhRvX}
---. 2016{\natexlab{b}}, Physical Review X, 6, 041015

\bibitem[{{Abbott} {et~al.}(2016{\natexlab{c}}){Abbott}, {Abbott}, {Abbott},
  {Abernathy}, {Acernese}, {Ackley}, {Adams}, {Adams}, {Addesso}, {Adhikari},
  \& et~al.}]{abbott2016PhRvL}
---. 2016{\natexlab{c}}, Physical Review Letters, 116, 061102

\bibitem[{{Antonini} \& {Rasio}(2016)}]{antonini2016ApJ}
{Antonini}, F., \& {Rasio}, F.~A. 2016, \apj, 831, 187

\bibitem[{{Barkov} \& {Komissarov}(2010)}]{barkov2010MNRAS}
{Barkov}, M.~V., \& {Komissarov}, S.~S. 2010, \mnras, 401, 1644

\bibitem[{{Bartos} {et~al.}(2017){Bartos}, {Kocsis}, {Haiman}, \&
  {M{\'a}rka}}]{bartos2017ApJ}
{Bartos}, I., {Kocsis}, B., {Haiman}, Z., \& {M{\'a}rka}, S. 2017, \apj, 835,
  165

\bibitem[{{Belczynski} {et~al.}(2016){Belczynski}, {Holz}, {Bulik}, \&
  {O'Shaughnessy}}]{belczynski2016Nature}
{Belczynski}, K., {Holz}, D.~E., {Bulik}, T., \& {O'Shaughnessy}, R. 2016,
  \nat, 534, 512

\bibitem[{{Bird} {et~al.}(2016){Bird}, {Cholis}, {Mu{\~n}oz},
  {Ali-Ha{\"i}moud}, {Kamionkowski}, {Kovetz}, {Raccanelli}, \&
  {Riess}}]{bird2016PhRvL}
{Bird}, S., {Cholis}, I., {Mu{\~n}oz}, J.~B., {Ali-Ha{\"i}moud}, Y.,
  {Kamionkowski}, M., {Kovetz}, E.~D., {Raccanelli}, A., \& {Riess}, A.~G.
  2016, Physical Review Letters, 116, 201301

\bibitem[{{Blandford} \& {Znajek}(1977)}]{brandford1977MNRAS}
{Blandford}, R.~D., \& {Znajek}, R.~L. 1977, \mnras, 179, 433

\bibitem[{{Blinnikov} {et~al.}(2016){Blinnikov}, {Dolgov}, {Porayko}, \&
  {Postnov}}]{blinnikov2016JCAP}
{Blinnikov}, S., {Dolgov}, A., {Porayko}, N.~K., \& {Postnov}, K. 2016, \jcap,
  11, 036

\bibitem[{{Bromberg} {et~al.}(2011){Bromberg}, {Nakar}, {Piran}, \&
  {Sari}}]{bromberg2011ApJ}
{Bromberg}, O., {Nakar}, E., {Piran}, T., \& {Sari}, R. 2011, \apj, 740, 100

\bibitem[{{Bromberg} {et~al.}(2012){Bromberg}, {Nakar}, {Piran}, \&
  {Sari}}]{bromberg2012ApJ}
---. 2012, \apj, 749, 110

\bibitem[{{Bulik} {et~al.}(2011){Bulik}, {Belczynski}, \&
  {Prestwich}}]{bulik2011ApJ}
{Bulik}, T., {Belczynski}, K., \& {Prestwich}, A. 2011, \apj, 730, 140

\bibitem[{{Carpano} {et~al.}(2007){Carpano}, {Pollock}, {Prestwich},
  {Crowther}, {Wilms}, {Yungelson}, \& {Ehle}}]{carpano2007A&A}
{Carpano}, S., {Pollock}, A.~M.~T., {Prestwich}, A., {Crowther}, P., {Wilms},
  J., {Yungelson}, L., \& {Ehle}, M. 2007, \aap, 466, L17

\bibitem[{{Crowther} {et~al.}(2010){Crowther}, {Barnard}, {Carpano}, {Clark},
  {Dhillon}, \& {Pollock}}]{crowther2010MNRAS}
{Crowther}, P.~A., {Barnard}, R., {Carpano}, S., {Clark}, J.~S., {Dhillon},
  V.~S., \& {Pollock}, A.~M.~T. 2010, \mnras, 403, L41

\bibitem[{{de Souza} {et~al.}(2011){de Souza}, {Yoshida}, \&
  {Ioka}}]{desouza2011A&A}
{de Souza}, R.~S., {Yoshida}, N., \& {Ioka}, K. 2011, \aap, 533, A32

\bibitem[{{Detmers} {et~al.}(2008){Detmers}, {Langer}, {Podsiadlowski}, \&
  {Izzard}}]{detmers2008A&A}
{Detmers}, R.~G., {Langer}, N., {Podsiadlowski}, P., \& {Izzard}, R.~G. 2008,
  \aap, 484, 831

\bibitem[{{Eggleton}(1983)}]{eggleton1983ApJ}
{Eggleton}, P.~P. 1983, \apj, 268, 368

\bibitem[{{Esposito} {et~al.}(2015){Esposito}, {Israel}, {Milisavljevic},
  {Mapelli}, {Zampieri}, {Sidoli}, {Fabbiano}, \& {Rodr{\'{\i}}guez
  Castillo}}]{esposito2015MNRAS}
{Esposito}, P., {Israel}, G.~L., {Milisavljevic}, D., {Mapelli}, M.,
  {Zampieri}, L., {Sidoli}, L., {Fabbiano}, G., \& {Rodr{\'{\i}}guez Castillo},
  G.~A. 2015, \mnras, 452, 1112

\bibitem[{{Farr} {et~al.}(2011){Farr}, {Sravan}, {Cantrell}, {Kreidberg},
  {Bailyn}, {Mandel}, \& {Kalogera}}]{farr2011ApJ}
{Farr}, W.~M., {Sravan}, N., {Cantrell}, A., {Kreidberg}, L., {Bailyn}, C.~D.,
  {Mandel}, I., \& {Kalogera}, V. 2011, \apj, 741, 103

\bibitem[{{Ghirlanda} {et~al.}(2016){Ghirlanda}, {Salafia}, {Pescalli},
  {Ghisellini}, {Salvaterra}, {Chassande-Mottin}, {Colpi}, {Nappo}, {D'Avanzo},
  {Melandri}, {Bernardini}, {Branchesi}, {Campana}, {Ciolfi}, {Covino},
  {G{\"o}tz}, {Vergani}, {Zennaro}, \& {Tagliaferri}}]{ghirlanda2016A&A}
{Ghirlanda}, G., {et~al.} 2016, \aap, 594, A84

\bibitem[{{Goldreich} \& {Nicholson}(1989)}]{goldreich1989ApJ}
{Goldreich}, P., \& {Nicholson}, P.~D. 1989, \apj, 342, 1079

\bibitem[{{Hirano} {et~al.}(2014){Hirano}, {Hosokawa}, {Yoshida}, {Umeda},
  {Omukai}, {Chiaki}, \& {Yorke}}]{hirano2014ApJ}
{Hirano}, S., {Hosokawa}, T., {Yoshida}, N., {Umeda}, H., {Omukai}, K.,
  {Chiaki}, G., \& {Yorke}, H.~W. 2014, \apj, 781, 60

\bibitem[{{Hirschi} {et~al.}(2004){Hirschi}, {Meynet}, \&
  {Maeder}}]{hirschi2004A&A}
{Hirschi}, R., {Meynet}, G., \& {Maeder}, A. 2004, \aap, 425, 649

\bibitem[{{Hosokawa} {et~al.}(2011){Hosokawa}, {Omukai}, {Yoshida}, \&
  {Yorke}}]{hosokawa2011Sci}
{Hosokawa}, T., {Omukai}, K., {Yoshida}, N., \& {Yorke}, H.~W. 2011, Science,
  334, 1250

\bibitem[{{Hurley} {et~al.}(2000){Hurley}, {Pols}, \& {Tout}}]{hurley2000MNRAS}
{Hurley}, J.~R., {Pols}, O.~R., \& {Tout}, C.~A. 2000, \mnras, 315, 543

\bibitem[{{Inayoshi} {et~al.}(2017){Inayoshi}, {Hirai}, {Kinugawa}, \&
  {Hotokezaka}}]{inayoshi2017}
{Inayoshi}, K., {Hirai}, R., {Kinugawa}, T., \& {Hotokezaka}, K. 2017, ArXiv
  e-prints

\bibitem[{{Ivanova} {et~al.}(2013){Ivanova}, {Justham}, {Chen}, {De Marco},
  {Fryer}, {Gaburov}, {Ge}, {Glebbeek}, {Han}, {Li}, {Lu}, {Marsh},
  {Podsiadlowski}, {Potter}, {Soker}, {Taam}, {Tauris}, {van den Heuvel}, \&
  {Webbink}}]{ivanova2013A&ARv}
{Ivanova}, N., {et~al.} 2013, \aapr, 21, 59

\bibitem[{{Janka}(2013)}]{janka2013MNRAS}
{Janka}, H.-T. 2013, \mnras, 434, 1355

\bibitem[{{Kashlinsky}(2016)}]{kashlinsky2016ApJ}
{Kashlinsky}, A. 2016, \apjl, 823, L25

\bibitem[{{Kinugawa} {et~al.}(2014){Kinugawa}, {Inayoshi}, {Hotokezaka},
  {Nakauchi}, \& {Nakamura}}]{kinugawa2014MNRAS}
{Kinugawa}, T., {Inayoshi}, K., {Hotokezaka}, K., {Nakauchi}, D., \&
  {Nakamura}, T. 2014, \mnras, 442, 2963

\bibitem[{{Kruckow} {et~al.}(2016){Kruckow}, {Tauris}, {Langer}, {Sz{\'e}csi},
  {Marchant}, \& {Podsiadlowski}}]{kruckow2016A&A}
{Kruckow}, M.~U., {Tauris}, T.~M., {Langer}, N., {Sz{\'e}csi}, D., {Marchant},
  P., \& {Podsiadlowski}, P. 2016, \aap, 596, A58

\bibitem[{{Kushnir} {et~al.}(2016){Kushnir}, {Zaldarriaga}, {Kollmeier}, \&
  {Waldman}}]{kushnir2016MNRAS}
{Kushnir}, D., {Zaldarriaga}, M., {Kollmeier}, J.~A., \& {Waldman}, R. 2016,
  \mnras, 462, 844

\bibitem[{{Kushnir} {et~al.}(2017){Kushnir}, {Zaldarriaga}, {Kollmeier}, \&
  {Waldman}}]{kushnir2017MNRAS}
---. 2017, \mnras, 467, 2146

\bibitem[{{Langer} {et~al.}(1994){Langer}, {Hamann}, {Lennon}, {Najarro},
  {Pauldrach}, \& {Puls}}]{langer1994A&A}
{Langer}, N., {Hamann}, W.-R., {Lennon}, M., {Najarro}, F., {Pauldrach},
  A.~W.~A., \& {Puls}, J. 1994, \aap, 290, 819

\bibitem[{{Liu} {et~al.}(2013){Liu}, {Bregman}, {Bai}, {Justham}, \&
  {Crowther}}]{liu2013Nature}
{Liu}, J.-F., {Bregman}, J.~N., {Bai}, Y., {Justham}, S., \& {Crowther}, P.
  2013, \nat, 503, 500

\bibitem[{{MacFadyen} \& {Woosley}(1999)}]{macfadyen1999ApJ}
{MacFadyen}, A.~I., \& {Woosley}, S.~E. 1999, \apj, 524, 262

\bibitem[{{Madau} \& {Dickinson}(2014)}]{madau2014ARA&A}
{Madau}, P., \& {Dickinson}, M. 2014, \araa, 52, 415

\bibitem[{{Mandel}(2016)}]{mandel2016MNRASb}
{Mandel}, I. 2016, \mnras, 456, 578

\bibitem[{{Mandel} \& {de Mink}(2016)}]{mandel2016MNRAS}
{Mandel}, I., \& {de Mink}, S.~E. 2016, \mnras, 458, 2634

\bibitem[{{Maoz} {et~al.}(2014){Maoz}, {Mannucci}, \&
  {Nelemans}}]{maoz2014ARA&A}
{Maoz}, D., {Mannucci}, F., \& {Nelemans}, G. 2014, \araa, 52, 107

\bibitem[{{Marchant} {et~al.}(2016){Marchant}, {Langer}, {Podsiadlowski},
  {Tauris}, \& {Moriya}}]{marchant2016}
{Marchant}, P., {Langer}, N., {Podsiadlowski}, P., {Tauris}, T.~M., \&
  {Moriya}, T.~J. 2016, \aap, 588, A50

\bibitem[{{Marigo} {et~al.}(2001){Marigo}, {Girardi}, {Chiosi}, \&
  {Wood}}]{marigo2001A&A}
{Marigo}, P., {Girardi}, L., {Chiosi}, C., \& {Wood}, P.~R. 2001, \aap, 371,
  152

\bibitem[{{Meynet} {et~al.}(2011){Meynet}, {Georgy}, {Hirschi}, {Maeder},
  {Massey}, {Przybilla}, \& {Nieva}}]{meynet2011BSRSL}
{Meynet}, G., {Georgy}, C., {Hirschi}, R., {Maeder}, A., {Massey}, P.,
  {Przybilla}, N., \& {Nieva}, M.-F. 2011, Bulletin de la Societe Royale des
  Sciences de Liege, 80, 266

\bibitem[{{Meynet} \& {Maeder}(2003)}]{meynet2003A&A}
{Meynet}, G., \& {Maeder}, A. 2003, \aap, 404, 975

\bibitem[{{Meynet} \& {Maeder}(2005)}]{maynet2005A&A}
---. 2005, \aap, 429, 581

\bibitem[{{Meynet} \& {Maeder}(2007)}]{meynet2007A&A}
---. 2007, \aap, 464, L11

\bibitem[{{Nakamura} {et~al.}(2016){Nakamura}, {Ando}, {Kinugawa}, {Nakano},
  {Eda}, {Sato}, {Musha}, {Akutsu}, {Tanaka}, {Seto}, {Kanda}, \&
  {Itoh}}]{nakamura2016PTEP}
{Nakamura}, T., {et~al.} 2016, Progress of Theoretical and Experimental
  Physics, 2016, 093E01

\bibitem[{{O'Connor} \& {Ott}(2011)}]{oconnor2011ApJ}
{O'Connor}, E., \& {Ott}, C.~D. 2011, \apj, 730, 70

\bibitem[{{O'Leary} {et~al.}(2016){O'Leary}, {Meiron}, \&
  {Kocsis}}]{oleary2016ApJ}
{O'Leary}, R.~M., {Meiron}, Y., \& {Kocsis}, B. 2016, \apjl, 824, L12

\bibitem[{{Omukai} \& {Palla}(2003)}]{omukai2003ApJ}
{Omukai}, K., \& {Palla}, F. 2003, \apj, 589, 677

\bibitem[{{{\"O}zel} {et~al.}(2010){{\"O}zel}, {Psaltis}, {Narayan}, \&
  {McClintock}}]{ozel2010ApJ}
{{\"O}zel}, F., {Psaltis}, D., {Narayan}, R., \& {McClintock}, J.~E. 2010,
  \apj, 725, 1918

\bibitem[{{Peters}(1964)}]{peters1964PhRv}
{Peters}, P.~C. 1964, Physical Review, 136, 1224

\bibitem[{{Podsiadlowski} {et~al.}(2004){Podsiadlowski}, {Mazzali}, {Nomoto},
  {Lazzati}, \& {Cappellaro}}]{podsiadlowski2004ApJ}
{Podsiadlowski}, P., {Mazzali}, P.~A., {Nomoto}, K., {Lazzati}, D., \&
  {Cappellaro}, E. 2004, \apjl, 607, L17

\bibitem[{{Prestwich} {et~al.}(2007){Prestwich}, {Kilgard}, {Crowther},
  {Carpano}, {Pollock}, {Zezas}, {Saar}, {Roberts}, \&
  {Ward}}]{pretwich2007ApJ}
{Prestwich}, A.~H., {et~al.} 2007, \apjl, 669, L21

\bibitem[{{Ram{\'{\i}}rez-Agudelo} {et~al.}(2015){Ram{\'{\i}}rez-Agudelo},
  {Sana}, {de Mink}, {H{\'e}nault-Brunet}, {de Koter}, {Langer}, {Tramper},
  {Gr{\"a}fener}, {Evans}, {Vink}, {Dufton}, \& {Taylor}}]{ram2015A&A}
{Ram{\'{\i}}rez-Agudelo}, O.~H., {et~al.} 2015, \aap, 580, A92

\bibitem[{{Rodriguez} {et~al.}(2016{\natexlab{a}}){Rodriguez}, {Haster},
  {Chatterjee}, {Kalogera}, \& {Rasio}}]{rodriguez2016ApJb}
{Rodriguez}, C.~L., {Haster}, C.-J., {Chatterjee}, S., {Kalogera}, V., \&
  {Rasio}, F.~A. 2016{\natexlab{a}}, \apjl, 824, L8

\bibitem[{{Rodriguez} {et~al.}(2016{\natexlab{b}}){Rodriguez}, {Zevin},
  {Pankow}, {Kalogera}, \& {Rasio}}]{rodriguez2016ApJ}
{Rodriguez}, C.~L., {Zevin}, M., {Pankow}, C., {Kalogera}, V., \& {Rasio},
  F.~A. 2016{\natexlab{b}}, \apjl, 832, L2

\bibitem[{{Sasaki} {et~al.}(2016){Sasaki}, {Suyama}, {Tanaka}, \&
  {Yokoyama}}]{sasaki2016PhRvL}
{Sasaki}, M., {Suyama}, T., {Tanaka}, T., \& {Yokoyama}, S. 2016, Physical
  Review Letters, 117, 061101

\bibitem[{{Sekiguchi} \& {Shibata}(2011)}]{sekiguchi2011ApJ}
{Sekiguchi}, Y., \& {Shibata}, M. 2011, \apj, 737, 6

\bibitem[{{Silverman} \& {Filippenko}(2008)}]{silverman2008ApJ}
{Silverman}, J.~M., \& {Filippenko}, A.~V. 2008, \apjl, 678, L17

\bibitem[{{Stone} {et~al.}(2017){Stone}, {Metzger}, \&
  {Haiman}}]{stone2017MNRAS}
{Stone}, N.~C., {Metzger}, B.~D., \& {Haiman}, Z. 2017, \mnras, 464, 946

\bibitem[{{Toledano} {et~al.}(2007){Toledano}, {Moreno}, {Koenigsberger},
  {Detmers}, \& {Langer}}]{toledano2007A&A}
{Toledano}, O., {Moreno}, E., {Koenigsberger}, G., {Detmers}, R., \& {Langer},
  N. 2007, \aap, 461, 1057

\bibitem[{{Tout} {et~al.}(1996){Tout}, {Pols}, {Eggleton}, \&
  {Han}}]{tout1996MNRAS}
{Tout}, C.~A., {Pols}, O.~R., {Eggleton}, P.~P., \& {Han}, Z. 1996, \mnras,
  281, 257

\bibitem[{{Ud-Doula} {et~al.}(2009){Ud-Doula}, {Owocki}, \&
  {Townsend}}]{Ud-Doula2009MNRAS}
{Ud-Doula}, A., {Owocki}, S.~P., \& {Townsend}, R.~H.~D. 2009, \mnras, 392,
  1022

\bibitem[{{van den Heuvel} {et~al.}(2017){van den Heuvel}, {Portegies Zwart},
  \& {de Mink}}]{vandenHeuvel2017}
{van den Heuvel}, E.~P.~J., {Portegies Zwart}, S.~F., \& {de Mink}, S.~E. 2017,
  ArXiv e-prints

\bibitem[{{Visbal} {et~al.}(2015){Visbal}, {Haiman}, \&
  {Bryan}}]{visbal2015MNRAS}
{Visbal}, E., {Haiman}, Z., \& {Bryan}, G.~L. 2015, \mnras, 453, 4456

\bibitem[{{Wanderman} \& {Piran}(2010)}]{wanderman2010MNRAS}
{Wanderman}, D., \& {Piran}, T. 2010, \mnras, 406, 1944

\bibitem[{{Witte} \& {Savonije}(1999)}]{witte1999A&A}
{Witte}, M.~G., \& {Savonije}, G.~J. 1999, \aap, 350, 129

\bibitem[{{Woosley}(1993)}]{woosley1993ApJ}
{Woosley}, S.~E. 1993, \apj, 405, 273

\bibitem[{{Woosley} \& {Heger}(2012)}]{woosley2012ApJ}
{Woosley}, S.~E., \& {Heger}, A. 2012, \apj, 752, 32

\bibitem[{{Zahn}(1975)}]{zahn1975A&A}
{Zahn}, J.-P. 1975, \aap, 41, 329

\bibitem[{{Zahn}(1977)}]{zahn1977A&A}
---. 1977, \aap, 57, 383

\bibitem[{{Zaldarriaga} {et~al.}(2017){Zaldarriaga}, {Kushnir}, \&
  {Kollmeier}}]{zaldarriaga2017}
{Zaldarriaga}, M., {Kushnir}, D., \& {Kollmeier}, J.~A. 2017, ArXiv e-prints

\end{thebibliography}


\end{document}